\documentclass[11pt]{article} 
\usepackage{cite}
\usepackage{latexsym}
\usepackage{amssymb}
\usepackage{amsmath}
\usepackage{bm}
\usepackage{hyperref}
\hypersetup{colorlinks=true, allcolors=blue}
\usepackage{graphicx}

\usepackage{slashed}
\usepackage{mathbbol}
\usepackage[title]{appendix}

\usepackage{color}

\begin{document}
\pagestyle{empty}

\begin{flushright}
KEK-TH-2311
\end{flushright}

\vspace{3cm}

\begin{center}

{\bf\LARGE  
Stochastic computation of $g-2$ in QED
}
\\

\vspace*{1.5cm}
{\large 
Ryuichiro Kitano$^{1,2}$, Hiromasa Takaura$^1$, and Shoji Hashimoto$^{1,2}$ 
} \\
\vspace*{0.5cm}

{\it 
$^1$KEK Theory Center, Tsukuba 305-0801,
Japan\\
$^2$Graduate University for Advanced Studies (Sokendai), Tsukuba
305-0801, Japan\\
}

\end{center}

\vspace*{1.0cm}

\begin{abstract}
{\normalsize
We perform a numerical computation of the anomalous magnetic moment
($g-2$) of the electron in QED by using the stochastic perturbation
theory. Formulating QED on the lattice, we develop a method to
calculate the coefficients of the perturbative series of $g-2$ without
the use of the Feynman diagrams. We demonstrate the feasibility of the
method by performing a computation up to the $\alpha^3$ order and
compare with the known results. This program provides us with a
totally independent check of the results obtained by the Feynman
diagrams and will be useful for the estimations of not-yet-calculated
higher order values. This work provides an example of the application
of the numerical stochastic perturbation theory to physical
quantities, for which the external states have to be taken on-shell.

}
\end{abstract} 

\newpage
\baselineskip=18pt
\setcounter{page}{2}
\pagestyle{plain}
\baselineskip=18pt
\pagestyle{plain}

\setcounter{footnote}{0}

\section{Introduction}

Feynman diagrams depict correlation functions in quantum field
theories as an expansion in terms of small coupling constants in the
Lagrangian.
They provide us with the prime tool to calculate physical quantities
once the expansion parameters are identified. Indeed, the most
successful prediction of the quantum field theory, the anomalous
magnetic moment ($g-2$) in QED, has been calculated up to
$O(\alpha^5)$ by the method of the Feynman diagrams (see
Ref.~\cite{Aoyama:2019ryr} for a review), and the value is confirmed
at the experiments with precisions of
$O(10^{-12})$~\cite{Hanneke:2008tm}!

The computation based on Feynman diagrams, however, becomes pretty
much complicated at higher orders. Already at $O(\alpha^2)$, it took
about a decade to converge on the correct
value~\cite{Petermann:1957hs, Sommerfield:1957zz} since the famous
leading-order formula, $g-2=\alpha/\pi$, was obtained by Schwinger in
1948~\cite{Schwinger:1948iu}. Even today, only experts can go beyond
two-loop. (See Ref.~\cite{Laporta:1996mq} for the analytic expression
at the three-loop level, which is obtained after a number of numerical
works~\cite{Cvitanovic:1974um, Kinoshita:1973ph, Levine:1974cb,
Carroll:1974bu, Carroll:1975jf, Kinoshita:1995ym}.)
Recently, Aoyama, Hayakawa, Kinoshita, and Nio have achieved the
calculation up to the five-loop level, which requires evaluations of
more than ten thousand diagrams~\cite{Aoyama:2008gy,
Aoyama:2008hz, Aoyama:2010pk, Aoyama:2010yt, Aoyama:2010zp,
Aoyama:2011dy, Aoyama:2011rm, Aoyama:2011zy, Aoyama:2012fc,
Aoyama:2012wj, Aoyama:2017uqe, Aoyama:2019ryr}, for which a numerical
method has been used for the integration over Feynman parameters. The
current numerical uncertainties of the highest order coefficient are
of order a few percent. An independent numerical calculation has been
performed for a subset of diagrams in Ref.~\cite{Volkov:2019phy}.
There is a discrepancy between two groups beyond the statistical
uncertainties although it is not a very serious problem in the current
experimental precisions.
It would not be feasible to go beyond five loops in near future due to
the explosion of the number of diagrams and also to the increase of
the dimension of the integrals. It is desired to have another method,
which does not use the Feynman diagrams, for higher order calculations
as well as for the check of the results up to five loops. Not only
limited to the $g-2$ computation, new methods to calculate the
perturbative series will be quite welcome in general.

There is indeed a simple numerical method to obtain the perturbative
series developed in Refs.~\cite{DiRenzo:1994av, DiRenzo:1994sy,
DiRenzo:2000qe, DiRenzo:2002vg} (see Ref.~\cite{DiRenzo:2004hhl} for a
review), which uses the stochastic quantization
method~\cite{Parisi:1980ys}. Correlation functions in the stochastic
quantization are evaluated as long-time averages of fictitious time
evolutions governed by the Langevin equation, and they are formally
equivalent to the results of the path integral quantization.
In the Langevin equation, one can expand the fields in terms of a
coupling constant, $\lambda$, such as $\phi = \phi^{(0)} + \lambda
\phi^{(1)} + \cdots$, and write down the evolution equations for each
$\phi^{(i)}$. The correlation functions at a fixed order
$O(\lambda^n)$ can be calculated as a sum of time averages, such as
$\displaystyle \sum_{i=0}^n \langle \phi^{(i)} \phi^{(n-i)} \rangle$
in the case of two-point functions.
There is no need of the Feynman diagrams in this method. All the
diagrams are automatically summed in solving the set of the Langevin
equations.
If one tries to obtain analytic results, however, we simply recover
calculations similar to the Feynman rules~\cite{Parisi:1980ys}.
The formalism is more suited for a numerical simulation.
By putting the theory on the lattice to make the degrees of freedom
finite, the direct integration of the Langevin equation becomes
possible. This provides us with an independent numerical method for
the evaluation of the coefficients of perturbation series.
Simplicity of the method is especially beneficial for higher-order
computations.

There have been recent works on the applications of the stochastic
perturbation theory in Yang-Mills theory and in QCD, where the
expectation values of the Wilson loop operator are calculated up to
$O(\alpha_S^{35})$~\cite{Bali:2014fea, DelDebbio:2018ftu}!
The calculation is quite similar to the usual Monte-Carlo simulations
of lattice QCD. The major differences are the expansion in terms of
the coupling constant and also the way to treat the fermions on the
lattice.
The Langevin equation contains the inverse of the Dirac operator,
which is the most time consuming part in the lattice simulation, while
in the case of perturbative calculations, the fast Fourier transform
significantly reduces the cost of the fermion
part~\cite{DiRenzo:2000qe, DiRenzo:2002vg}. This makes it possible to
evaluate the coefficients at very high orders in QCD with dynamical
fermions.
The behavior of the expectation value at high orders has been used to
extract the renormalon contributions~\cite{Bali:2014sja,
DelDebbio:2018ftu}.

In this paper, we perform a QED computation of the electron $g-2$ in
the numerical stochastic perturbation theory. Since the $g-2$ factor
in perturbative QED is one of the simplest physical quantities (and
thus scheme independent) in particle physics, it is a good testing
ground for the formalism. Also, if it is successful, one can provide a
non-trivial check of the diagrammatic calculations as well as the
predictions of higher order coefficients.
We construct the action of lattice QED suitable for the stochastic
calculation, and perform the $g-2$ evaluation up to $O(\alpha^3)$ on
small lattices. 
The results are encouraging and we expect that simulations in
currently available large scale computers and/or more optimized
simulation codes can confirm the $O(\alpha^5)$ results of the
diagrammatic methods.

Matching to the continuum theory requires various extrapolations such
as the large volume limit, the continuum limit, as well as the limit
to the on-shell fermion momentum. Among them, the on-shell
extrapolation needs some indirect method as we work in the Euclidean
space.
In this work, we scan the fermion energies in the Euclidean region,
and infer on-shell values by using analytic continuation. 
We discuss how such extrapolations cause uncertainties in the
estimates of $g-2$. We also discuss infrared (IR) problems in QED in a
finite volume.

\section{Lattice QED action}

We first define the action of lattice QED which is to be used in the
simulation. Unlike the non-perturbative lattice simulation, the
fermion mass, $m_f$, is the only dimensionful parameter in question as
the Landau pole does not show up in the perturbative fixed-order
calculations. The continuum limit is, therefore, obtained by taking
the dimensionless combination $m_f a$ to be zero while fixing the pole
mass, $m_f$, to be finite, where $a$ is the lattice spacing.
For this purpose, it is desirable to maintain the chiral symmetry on
the lattice so that the fermion mass parameter $m$ in the Lagrangian
is not additively renormalized. If $m \propto m_f$, one can simply
take $ma \to 0$ as the continuum limit.

In the practical calculation, we need an IR regulator to eliminate the
zero mode of the gauge field to avoid its numerically unstable
random-walk evolution under the Langevin equation. We also need a
gauge fixing term to eliminate the gauge degrees of freedom.
Although the ultraviolet (UV) part is already regulated by the
lattice, we introduce a UV cut-off to avoid large logarithmic
corrections, which makes the continuum limit far away.

Considering those requirements, we define the action in the lattice
unit ($a=1$) as follows:
\begin{align}
 S_{\rm lattice} = S_{\rm g} + S_{\rm gf} + S_{\rm mass} + S_{\rm f},
 \label{eq:action}
\end{align}
where
\begin{align}
  S_{\rm g} = {1 \over 4}
  \sum_{n, \mu, \nu}
  \left[
    e^{-\nabla^2 / \Lambda_{\rm UV}^2}
    \left(
    \nabla_\mu A_\nu (n) - \nabla_\nu A_\mu (n)
    \right)
    \right]^2,
  \label{eq:action_mod1}
\end{align}
\begin{align}
  S_{\rm gf} = {1 \over 2 \xi} \sum_{n} \left[
    e^{-\nabla^2 / \Lambda_{\rm UV}^2} \sum_\mu \nabla_\mu^* A_\mu (n)
    \right]^2,
  \label{eq:action_mod2}
\end{align}
\begin{align}
  S_{\rm mass} = {1 \over 2} \sum_{n, \mu} m_{\gamma}^2
  \left[
    e^{-\nabla^2 / \Lambda_{\rm UV}^2} A_\mu (n)
    \right]^2,
  \label{eq:action_mod3}
\end{align}
and
\begin{align}
S_{\rm f} = - {1 \over 16} \ln {\rm det} D.
\label{eq:action_fermion}
\end{align}
The derivatives are defined by
\begin{align}
 \nabla_\mu f (n) = f(n+ \hat \mu) - f(n), \quad
 \nabla_\mu^* f (n) = f(n) - f(n - \hat \mu),
\end{align}
and $\nabla^2 = \nabla_\mu^* \nabla_\mu$.
The Dirac operator $D$ is that of the naive fermion,
\begin{align}
 (D)_{nm} = m \delta_{nm}
+ {1 \over 2} \sum_\mu \left[
\gamma_\mu e^{i e A_\mu (n)} \delta_{n+\hat \mu,m}
- \gamma_\mu e^{-i e A_\mu (n-\hat \mu)} \delta_{n-\hat \mu,m}
\right],
\label{eq:dirac_operator}
\end{align}
with $m$ the fermion mass parameter.
The chiral symmetry is maintained on the lattice because we use the
naive Dirac operator, with which the fermion propagator has sixteen
poles. Those doublers are taken care by the factor of $1/16$ in
Eq.~\eqref{eq:action_fermion}. By this treatment, the action describes
the system with a single fermion at least for perturbative
calculations of quantities which do not involve $\gamma_5$ in the
internal line. For the external fermion lines, one can choose the
external momenta such that the doubler poles are not important.

The terms $S_{\rm g}$ and $S_{\rm f}$ are invariant under the gauge
transformation:
\begin{align}
 A_\mu (n) = A_\mu (n) + {1 \over e} \nabla_\mu \theta (n).
\end{align}
The gauge fixing term $S_{\rm gf}$ eliminates the gauge degrees of
freedom~\cite{Zwanziger:1981kg}.
We take the non-compact QED action $S_{\rm g}$ so that there is no
self interactions of photons. This simplifies the perturbative
expansion of the Langevin equation compared to the case of QCD.
The QED coupling constant $e$ only appears in
Eq.~\eqref{eq:dirac_operator}.

A UV regulator is introduced through the factor $e^{-\nabla^2 /
\Lambda_{\rm UV}^2}$ in the gauge action $S_{\rm g} + S_{\rm gf} +
S_{\rm mass}$. This factor reduces the strength of the gauge
interaction at high photon virtuality $\hat k^2$ in a gauge invariant
way. By an appropriate choice of $\Lambda_{\rm UV}$, one can reduce
large logarithmic corrections which appear in the renormalization
factors.
In the continuum limit, we should send the UV cut-off to infinity
while the physical fermion mass $m_f \propto m$ fixed. This means we
need to take the dimensionless quantity $ma$ sufficiently small, while
$\Lambda_{\rm UV} a$ is fixed. Therefore, in the actual simulation, we
can simply use a fixed value of $\Lambda_{\rm UV}$ in the $a=1$ unit.
In the numerical calculation, this UV regularization is quite cheaply
implemented by the use of the fast Fourier transform: going to the
momentum space, multiply $e^{\hat k^2 / \Lambda_{\rm UV}^2}$ and
coming back to the position space.

We introduce the photon mass $m_\gamma$ as the IR regulator as we
usually do in the perturbative calculations in the on-shell scheme. We
should take the limit of $m_\gamma/m \to 0$ later. 
The gauge invariance is broken by this procedure, but there are
modified Ward-Takahashi identities that restrict the form of the
quantum corrections. For detail, see Appendix~\ref{sec:WT}.
Since the zero mode is regulated by the mass term, one can simply take
the periodic boundary conditions for the gauge field. 

In the literature, different IR regularizations in the lattice
perturbation theories have been discussed. One is to remove the zero
mode from the summation of the momentum~\cite{Heller:1984hx,
Hayakawa:2008an}. Although this would be the simplest implementation,
the estimation of the finite volume effects is quite non-trivial after
the removal of the zero mode as discussed in
Ref.~\cite{Davoudi:2018qpl}. When the zero mode is just removed, the
theory is no longer a local field theory.
Since we will heavily use the analyticity of the correlation functions
in later discussions, non-locality of the theory may introduce an
additional complication to the argument.

Imposing the twisted boundary condition is another way to remove the
zero mode from the gauge field in the case of $SU(N)$ gauge
theories~\cite{Luscher:1985wf}. In $U(1)$ gauge theories, a similar
regularization may be possible by introducing a background magnetic
flux on one of the two tori, $\int_{T^2} dA = 2\pi/e$.
However, since we use the non-compact QED action, where $A_\mu (n)$ is
treated as the fundamental field, there cannot be a magnetic monopole
background to realize the flux.

For these reasons, we employ the simple IR regularization by the
photon mass term. The effects due to the mass term are studied in
Refs.~\cite{Endres:2015gda, Endres:2015vpi}, and it is found that the
zero mode does modify the analytic structure, but in a controlled way.
We will discuss it in Section~\ref{sec:9}.

\section{Langevin equation}
\label{sec:langevin}

In the stochastic quantization, one can obtain the expectation values
of operator products as a ``time'' average of the field products. The
gauge field $A_\mu (n)$ is promoted to obtain another ``time''
coordinate $\tau$, $A_\mu (n, \tau)$, and the evolution in the $\tau$
direction is governed by the Langevin equation:
\begin{align}
 {\partial A_\mu (n,\tau) \over \partial \tau}
  = - {\delta S_{\rm lattice} \over \delta A_\mu (n,\tau)} + \eta_\mu (n,\tau).
 \label{eq:langevin}
\end{align}
The Gaussian noise $\eta_\mu (n,\tau)$ satisfies:
\begin{align}
 \langle \eta_\mu (n,\tau) \rangle_\eta = 0, \quad
 \langle \eta_\mu (n,\tau) \eta_\nu (n',\tau') \rangle_\eta
  = 2 \delta_{\mu \nu} \delta_{n n'}  \delta (\tau-\tau'),
\end{align}
where $\langle \cdots \rangle_\eta$ denotes an ensemble average over
the random noise.
The ensemble average of field products converges to the correlation
functions obtained by the path integral quantization:
\begin{align}
 \langle A_{\mu_1} (n_1, \tau) \cdots A_{\mu_k} (n_k, \tau) \rangle_\eta \to
 \langle A_{\mu_1} (n_1) \cdots A_{\mu_k} (n_k) \rangle,
\end{align}
as $\tau \to \infty$. This can be understood by the fact that the
probability distribution of the path integral, $e^{-S}$, is the fixed
point of the Fokker-Planck equation derived from the Langevin equation
in Eq.~\eqref{eq:langevin}.
The ensemble average in the left-hand-side can be evaluated as the
``time'' average of the Langevin trajectories of the field product, i.e.,
\begin{align}
  \langle A_{\mu_1} (n_1) \cdots A_{\mu_k} (n_k) \rangle 
  = \lim_{\Delta \tau \to \infty} {1 \over \Delta \tau}
  \int_{\tau_0}^{\tau_0 + \Delta \tau} d\tau 
  A_{\mu_1} (n_1, \tau) \cdots A_{\mu_k} (n_k, \tau).
 \end{align}
Therefore, by following long enough Langevin trajectories of fields,
one can calculate the correlation functions. For a numerical method to
integrate the Langevin equation, see
Appendix~\ref{sec:numerical_langevin}.

For our QED action, the only non-trivial part in the right-hand-side
of Eq.~\eqref{eq:langevin} is the fermion part:
\begin{align}
 {\delta S_{\rm f} \over \delta A_\mu (n)}
= - {1 \over 16} {\rm Tr} \left(
{\delta D \over \delta A_\mu (n)} D^{-1}
\right).
\label{eq:fermion-part}
\end{align}
The inverse of the Dirac operator, $D^{-1}$, is evaluated
perturbatively as we see later.
The trace of the space-time and spinor indices are taken by using a
noise spinor $\zeta$ as:
\begin{align}
 {\rm Tr} \left(
{\delta D \over \delta A_\mu (n)} D^{-1}
 \right) =
 \left\langle \zeta^\dagger {\delta D \over \delta A_\mu (n)} D^{-1}
 \zeta
 \right\rangle_\zeta,
 \label{eq:noise}
\end{align}
where the noise spinor satisfies
\begin{align}
 \langle \zeta(n)_\alpha \zeta(m)^\dagger_\beta \rangle_{\zeta} 
 = \delta_{nm} \delta_{\alpha \beta}.
\end{align}
The indices $\alpha$ and $\beta$ denote those of the spinor.
In practice, we generate only one noise spinor $\zeta$ for each
discretized time step of the numerical Langevin simulation. The
average is, therefore, taken during the large time evolution.

\section{Perturbative expansion}
\label{sec:perturbation}

We decompose the Langevin equation into those for each order in the
perturbative expansion. We follow the formalism of
Refs.~\cite{DiRenzo:2000qe, DiRenzo:2004hhl} for the treatment of the
fermions.

We expand the field $A_\mu (n,\tau)$ as
\begin{align}
 A_\mu (n, \tau) = \sum_{p=0}^{\infty} e^p A_\mu^{(p)} (n,\tau),
 \label{eq:expansion}
\end{align}
and solve the Langevin equation for each $A_\mu^{(p)}$.
The equation for gauge fields in each order $p$ is given by
\begin{align}
{\partial A_\mu^{(p)} (n,\tau) \over \partial \tau} 
= - {\delta S_{\rm lattice} \over \delta A_\mu (n,\tau)} \Bigg|_{(p)}
  + \eta_\mu (n, \tau) \delta_{p0},
\end{align}
where $|_{(p)}$ denotes the $p$-th order term. The noise is applied
only for the lowest order, $A_\mu^{(0)}$. Higher order fields get
fluctuated through interactions with lower order fields.

The expansion of the force term, $\delta S / \delta A_\mu$, is
straightforward except for the fermion part, for which
\begin{align}
 {\delta S_{\rm f} \over \delta A_\mu (n)}   \Bigg|_{(p)}
&= - {\rm Tr} \left(
{\delta D \over \delta A_\mu (n)} D^{-1}
\right)   \Bigg|_{(p)}
\nonumber \\
&= - \sum_{q=0}^{p} {\rm Tr} \left(
{\delta D \over \delta A_\mu (n)}  \Bigg|_{(p-q)} D^{-1}  \Big|_{(q)}
\right).
\end{align}
The vertex factor $\delta D / \delta A_\mu$ is evaluated as
\begin{align}
\left(
{\delta D \over \delta A_\mu (n)} 
\right)_{lm}    \Bigg|_{(0)}
= 0,
\end{align}
\begin{align}
\left(
{\delta D \over \delta A_\mu (n)} 
\right)_{lm} \Bigg|_{(p)}
=  {i \over 2} \left[
 \gamma_\mu U_{n,\mu}^{(p-1)} \delta_{n+\hat \mu, m} \delta_{nl}
+
 \gamma_\mu (U_{n,\mu}^{(p-1)})^* \delta_{n, m} \delta_{l-\hat \mu, n}
\right].
\end{align}
where
\begin{align}
U_{n,\mu}^{(p)}
&=e^{i e A_\mu (n)} \Big|_{(p)}.
\end{align}

The inverse of the Dirac operator, $D^{-1}$, can be obtained
recursively~\cite{DiRenzo:2000qe} as
\begin{align}
 (D^{-1}) \Big|_{(0)} = D_0^{-1}, \quad
  (D^{-1}) \Big|_{(p)} = - \left[ 
    D_0^{-1} D  \sum_{q=0}^{p-1} e^q
 \left( D^{-1} \right) \Big|_{(q)}
 \right] \Bigg|_{(p)}.
 \label{eq:inverse}
\end{align} 
Here, the propagator at the lowest order, $D_0^{-1}$, is given explicitly by
\begin{align}
\left( D_0^{-1} \right)_{nm}
&= \int {d^4 p \over (2 \pi)^4}
{-i \slashed s (p) + m
\over
s(p)^2 + m^2} e^{i p \cdot (x_n - x_m)},
\end{align}
where
\begin{align}
 s_\mu (p) = \sin (p_\mu).
\end{align}
The momentum integration should be understood as
\begin{align}
  \int {d p \over 2 \pi} \to {1 \over L} \sum_{k=0}^{L-1}, \label{eq:int_sum}
\end{align}
with $p = 2 \pi k / L$ for the periodic boundary conditions.

The propagator $D_0^{-1}$ is diagonal in the momentum space, while $D$
is local in the position space.
Therefore, the multiplication of the Dirac operators in
Eq.~\eqref{eq:inverse} can be effectively performed by the use of the
fast Fourier transform in the numerical calculation. This
significantly reduces the computational cost of the fermion
implementation in the stochastic perturbation theory.

In the calculation, all the fields as well as their correlation
functions are expanded in terms of the coupling constant $e$ as in
Eq.~\eqref{eq:expansion}. The operations of the expanded quantities,
such as the multiplication and the exponentiation are performed
perturbatively as
\begin{align}
  {\cal O}_1 {\cal O}_2 = \sum_p e^p 
  \sum_{q=0}^{p} {\cal O}_1^{(q)} {\cal O}_2^{(p-q)},
  \label{eq:product}
\end{align}
\begin{align}
  \exp (i e {\cal O}) = \sum_p {(ie)^p {\cal O}^p \over p!},
\end{align}
where ${\cal O}^p$ is defined by repeatedly using
Eq.~\eqref{eq:product}. General functions such as the inverse and
the square root of some operator are defined by the Taylor expansion in $e$,
where the lowest order value, ${\cal O}^{(0)}$, is formally assumed to be non-zero.

\section{Correlation functions}
\label{sec:correlators}

For the calculation of $g-2$, we first compute the following set of
correlation functions by using the Langevin evolutions.
In the following discussion, all the correlation functions are
implicitly expanded in terms of $e$. The parameters in the Lagrangian,
$ma$, $m_\gamma a$, $\Lambda_{\rm UV} a$, and $\xi$ on the other hand,
are kept as constants of $O(e^0)$.
In practice, we store the gauge field and the correlation functions as
arrays of perturbative coefficients, and their operations such as
multiplication are performed by applying the rule in
Eq.~\eqref{eq:product}.

\subsection{Photon two-point function}

The two-point function of the gauge field in the momentum space has
the form
\begin{align}
\langle \tilde A_\mu (k) \tilde A_\nu (k') \rangle
 = &
 e^{- 2 \hat k^2 / \Lambda_{\rm UV}^2} \left[
 {  \left(
 g_{\mu \nu} - {\hat k_\mu \hat k_\nu \over \hat k^2}
  \right)} {1 \over \hat k^2 + m_\gamma^2 - \Pi(\hat k^2)}
 + {\hat k_\mu \hat k_\nu \over \hat k^2}
 {\xi \over \hat k^2 + \xi m_\gamma^2}
 \right]\nonumber \\
 & \times
(2 \pi)^4 \delta^4 (k + k'),
\label{eq:AA}
\end{align}
where $\hat k_\mu = 2 \sin (k_\mu /2 )$, and $\hat k^2 = \hat k_\mu
\hat k_\mu$. The projection operators, $g_{\mu \nu} - \hat k_\mu \hat
k_\nu / \hat k^2$ and $\hat k_\mu \hat k_\nu / \hat k^2$, are
ambiguous at $\hat k^2 = 0$, but it is not important in the following
discussion, because we only use the two-point function at finite $\hat
k^2$.
The Fourier modes are defined as
\begin{align}
 A_\mu (n) = \int {d^4 k \over (2 \pi)^4} \tilde A_\mu (k) e^{i k \cdot (x_n + \hat \mu / 2)}, \quad
 \tilde A_\mu (k) = \sum_n A_\mu (n) e^{- i k \cdot (x_n + \hat \mu / 2)},
\end{align}
where $x_n + \hat \mu / 2$ stands for the mid point of the link specified
by the site $x_n$ and the direction $\mu$.
The vacuum polarization, $\Pi(\hat k^2)$, is vanishing at the tree
level, and is implicitly expanded as the perturbative series through
the expansion of the two-point function in Eq.~\eqref{eq:AA}.
The longitudinal part has no quantum correction as it is guaranteed by
the Ward-Takahashi identity (see Appendix~\ref{sec:WT}):
\begin{align}
\hat k_\mu \langle \tilde A_\mu (k) \tilde A_\nu (q) \rangle
= e^{-2 \hat k^2 / \Lambda_{\rm UV}^2} 
{\xi \hat k_\nu \over \hat k^2 + \xi m_\gamma^2} 
(2 \pi)^4 \delta^4 (k + q).
\end{align}
This fact is useful for the check of the gauge invariance.

We define the renormalization factor $Z_3$ as
\begin{align}
 Z_3^{-1} (\hat k^2) & = e^{2 \hat k^2 / \Lambda_{\rm UV}^2} \left(
   1 - {\Pi (\hat k^2) \over \hat k^2}
   \right).
 \label{eq:z3def}
\end{align}
By this $Z_3$ factor, one can obtain the physical coupling constant,
$e_{\rm P}$, as a power series of $e$ as follows:
\begin{align}
  e_{\rm P} = Z_3^{1/2} (0) e,
  \label{eq:eP}
\end{align}
where the limit of $m_\gamma^2 \to 0$ should be taken before $\hat k^2
\to 0$. This coupling constant, $e_{\rm P}$, characterizes the
coefficient of the long-range Coulomb force, and thus it is identified
as the fine-structure constant
\begin{align}
  \alpha = {e_{\rm P}^2 \over 4 \pi}
\end{align}
in the quantum mechanics.

We define the full photon propagator as
\begin{align}
 {\cal D}_{\mu \nu} (k) \equiv {1 \over V} 
 \langle \tilde A_\mu (k) \tilde A_\nu (-k) \rangle,
 \label{eq:photonprop}
\end{align}
where $V$ is the space-time volume, $L^3 \times T$.

\subsection{Fermion two-point function}

The two-point function of the fermions is computed as 
 \begin{align}
\left \langle
  \tilde D^{-1} (p,q)
  \right \rangle
  = \sum_{n,m} \left \langle
 (D^{-1})_{nm}
\right \rangle  e^{-i p \cdot x_n} e^{-i q \cdot x_m},
 \end{align}
where $\tilde D (p,q)$ is the Dirac operator in the momentum space.
The propagator should have poles at the on-shell energy:
\begin{align}
  \left \langle
  \tilde D^{-1} (p,q)
  \right \rangle
 \bigg|_{s(p)^2 \to - m_f^2}
\to Z_2 {-i \slashed s(p) + m_f \over s(p)^2 + m_f^2} (2 \pi)^4 \delta^4 (p + q),
\label{eq:near_on_shell}
\end{align}
where $Z_2$ denotes the renormalization factor for the fermion wave
function. Doubler poles are included as we do not try to eliminate
them in the action. We define the full fermion propagator as
\begin{align}
S (p) \equiv
{1 \over V} \left \langle
 \tilde D^{-1} (p,-p)
\right \rangle.
\label{eq:sf}
\end{align}

\subsection{Three-point function}

We are interested in the three-point function in the momentum space,
\begin{align}
G_\mu (p,k) 
&= {1 \over V} \sum_{n,m,l} \langle \psi (n) \bar \psi (m) A_\mu (l) 
\rangle e^{-i p \cdot x_n} e^{-i (-p - k) \cdot x_m} e^{- i k \cdot (x_l + \mu /2 )},
\nonumber \\
&= {1 \over V} \langle \tilde D^{-1}(p,-p-k) \tilde A_\mu (k) \rangle
\label{eq:gmu}
.
\end{align}
The photon momentum is denoted as $k$, and two external fermions have
the momentum $p$ and $p+k$, respectively.
The three-point function includes diagrams where the external photon
is not attached to the line of the external fermions (in the language
of Feynman diagrams). Among such diagrams, one-particle
irreducible ones (the light-by-light scattering) start to appear from
the $e^7$ order, {\it i.e.}, at the three-loop level. Since we are
interested in the three-loops and higher coefficients as well, we use
the three-point function defined above rather than $\langle \tilde
D^{-1} \gamma_\mu \tilde D^{-1} \rangle$, which is commonly used in the
lattice calculations of flavor non-singlet form factors.

The effective vertex function can be defined by amputating the
external legs as
\begin{align}
  - i e_{\rm P} \Gamma_\mu (p, k)
  =     &
  \kappa {\cal D}^{-1}_{\mu \nu} (k)
  S(p)^{-1} G_\nu (p, k)
  S(p+k)^{-1}.
  \label{eq:Gamma}
\end{align}
The prefactor $\kappa$ is a product of the appropriate wave-function
renormalization factors, which are not needed to be specified in the
following discussion.
For on-shell fermions, $s(p)^2 + m_f^2 = 0$ and $s(p+k)^2 + m_f^2 =
0$, the vertex function can be expressed by form factors as 
\begin{align}
-i e_{\rm P} \bar u(p) \Gamma_\mu (p,k) u(p+k)
 &=
-i e_{\rm P}  \bar u(p) \left(
F_1 (\hat k^2) \gamma_\mu
  - F_2 (\hat k^2)
 {\sigma_{\mu \nu} \hat k_\nu \over 2m_f}
 \right) u(p+k) +\mathcal{O}(a^2) ,
 \label{eq:form_factor}
\end{align}
where $u(p)$ and $u(p+k)$ are fermion wave functions, and $\sigma_{\mu
\nu} = (i / 2)[\gamma_\mu, \gamma_\nu]$. We used the equation of
motion to obtain this formula.
On the Euclidean lattice, one cannot directly go to the on-shell
momentum because of $s(p)^2 \geq 0$. We will discuss how on-shell
values can be extracted in the next section.

In the above expression, the $\mathcal{O}(a^2)$ correction contains
the violation of the Lorentz invariance on the lattice. At the tree
level, the form factor $F_1$ has the form $F_1^{\rm tree} = \cos
(p_\mu + k_\mu /2)$ in the unit of $a=1$ and thus depends on the
direction. In the continuum limit, it is reduced to $F_1^{\rm tree}(0)
= 1$ and thus the Lorentz invariance is recovered.

The physical QED coupling, $e_{\rm P}$, is defined such that
\begin{align}
 F_1 (0) = 1
\label{eq:onshell_renormalization}
\end{align}
in the continuum limit to all orders in perturbation theory. This
should coincide with the definition in terms of the long-range Coulomb
force in Eq.~\eqref{eq:eP}.

The $g$-factor to represent the magnetic moment is given by
\begin{align}
 {g \over 2} = {F_1 (0) + F_2 (0) \over F_1 (0)}.
 \label{eq:gfactor_def}
\end{align}
Our goal is to express this quantity as a power series of the
renormalized coupling $e_{\rm P}$.
At the tree level, $F_2(0) = 0$, and thus it seems automatic to obtain
$g=2$. However, as we will see later, we evaluate the numerator and
the denominator separately by projecting $\Gamma_\mu$ in the spatial
and the time directions, respectively.
Since $F_1(0)$ depends on the direction as already noted, we do not
obtain $g=2$ exactly at finite lattice spacings in our calculation.

The Ward identity:
\begin{align}
 - i e_{\rm P}  \hat k_\mu \Gamma_\mu (p,k) \cdot \kappa^{-1} 
= - e 
\left(
S^{-1} (p+k) - S^{-1} (p)
 \right)
 \label{eq:WardIdentity}
\end{align}
holds for non-vanishing photon masses. (See
Appendix~\ref{sec:WT}.) 
Noting Eq.~\eqref{eq:Gamma}, we can calculate the left-hand side as
the product of the correlation functions and there is no need to know
$\kappa$ or $e_{\rm P}$.
This relation is useful for the check of the gauge invariance in the
numerical calculations.

\section{Magnetic moments}
\label{sec:g-2}

We discuss how the ratio of the form factors at the on-shell fermion
momenta are extracted from the correlation functions calculated on the
Euclidean lattice.
In the following discussion, we take a particular configuration of the
external momenta ${\bf p} = -{\bf k}/2$ and $k_4 = 0$. The fermion
energy $p_4$ is left as a variable.
Since we take the ratio in Eq.~\eqref{eq:gfactor_def}, we do not need
to evaluate the normalization factor, $\kappa$, in
Eq.~\eqref{eq:Gamma}.
We first remove the photon propagator from the three-point function
$G_\mu$ in Eq.~\eqref{eq:gmu} by
\begin{align}
\hat G_\mu = 
{\cal D}^{-1}_{\mu \nu} (k)
G_\nu (p, k).
\label{eq:photon_removal}
\end{align}
We also define
\begin{align}
\hat G_\mu^{\rm (norm)} = -i S (p) \gamma_\mu S (p+k),
\label{eq:Gnorm}
\end{align}
by using the fermion two-point functions.
Since we choose $p_4 = p_4 + k_4$ and ${\bf p} = - ({\bf p} + {\bf
k})$, the two-point functions have a relation,
\begin{align}
  S(p) = \gamma_4 S(p+k) \gamma_4,
  \label{eq:SSrelation}
\end{align}
by parity transformation. Therefore, in the numerical calculation, we
only need to evaluate $\tilde D^{-1} (p', -p-k)$, which is obtained by
the multiplication of $\tilde D^{-1}$ to the source spinor at $p+k$ in
the momentum space. The propagator $S(p+k)$ is extracted as the
component at $p' = p+k$, and the above relation gives $S(p)$,
simultaneously.
The three-point function in Eq.~\eqref{eq:gmu} is obtained as the
$p'=p$ component multiplied by $\tilde A_\mu (k)$.

We scan all possible values of $p_4$ on the lattice, and calculate the
Fourier transform of appropriately projected functions:
\begin{align}
{\cal F}_E (t) = \int {d p_4 \over 2 \pi} {\rm tr} \left[
\gamma_4 \hat G_4
\right] e^{i p_4 t},
\label{eq:fE}
\end{align}
and
\begin{align}
{\cal F}_M (t)=  \int {d p_4 \over 2 \pi} \sum_{i,j,k = 1}^{3} i \epsilon_{ijk}
 {\rm tr} \left[ \gamma_5 \gamma_i \hat G_j \right]
\hat k_k  e^{i p_4 t}.
\label{eq:fM}
\end{align}
We note that the above integrals are understood as the abbreviation of
the loop sums [cf.~Eq.~\eqref{eq:int_sum}]. We repeat the same
calculation for $\hat G_\mu^{\rm (norm)}$ to obtain ${\cal F}_E^{\rm
(norm)} (t)$ and ${\cal F}_M^{\rm (norm)} (t)$. By defining a function
$g(t)$ as
\begin{align}
  {g (t) \over 2} = {
    {{\cal F}_M (t) / {\cal F}_E (t)} \over
    {{\cal F}_M^{\rm (norm)} (t) / {\cal F}_E^{\rm (norm)} (t)}
  },
  \label{eq:gfactor_formula}
\end{align}
the $g$ factor is obtained from $g(t)$ in the limit of $t\to \infty$ and
$\hat k \to 0$.

One can understand this formula as follows. First, there are useful
formulae,
\begin{align}
{\rm tr}& \left[
 P_{\rm E} \left(
F_1 (\hat k^2) \gamma_4 - F_2 (\hat k^2) {\sigma_{4 \mu} \hat
 k_{\mu} \over 2 m_f}
 \right)
 \right]
 = F_1 (\hat k^2) - {\hat {\bf k}^2 \over 4 m_f^2} F_2 (\hat k^2),
\end{align}
\begin{align}
\sum_{i,j,k = 1}^{3} \epsilon_{ijk} {\rm tr}& \left[
 (P_{\rm M})_i \left(
F_1 (\hat k^2) \gamma_j - F_2 (\hat k^2) {\sigma_{j \mu} \hat
 k_{\mu} \over 2 m_f}
 \right)
 \right] \hat k_k
 = F_1 (\hat k^2) + F_2 (\hat k^2),
\end{align}
where the projection operators, $P_E$ and $P_M$, are defined as
\begin{align}
 P_{\rm E} & = 
{1 \over 8 m_f^2}
 \left(
 - i \gamma_4 (iE) -i {\bm \gamma} \cdot {\bm s} (p + k)  + m_f
 \right) \gamma_4
\left(
 - i  \gamma_4 (iE) -i {\bm \gamma} \cdot {\bm s} (p)  + m_f 
 \right),
 \label{eq:pE}
\end{align}
and
\begin{align}
 ( P_{\rm M} )_\mu & = 
{i \over 8 \hat {\bf k}^2 E}
 \left(
 - i \gamma_4 (iE) -i {\bm \gamma} \cdot {\bm s} (p + k)  + m_f
 \right) \gamma_5 \gamma_\mu
\left(
 - i \gamma_4 (iE) -i {\bm \gamma} \cdot {\bm s} (p)  + m_f 
 \right).
 \label{eq:pM}
\end{align}
Here $E$ and $m_f$ are the fermion on-shell energy and the pole mass,
respectively, that satisfy
\begin{align}
  E^2 = m_f^2 + \sum_i s_i (p)^2 = m_f^2 + \sum_i s_i (p+k)^2.
\end{align}
Under the projection operators, the equation of motion can be used,
and thus we obtain
\begin{align}
 F_1 (\hat k^2)
 - {\hat {\bf k}^2 \over 4 m_f^2} F_2 (\hat k^2)&
 =  {\rm tr} \left[ P_{\rm E} \Gamma_4 \right] \bigg|_{\rm on\mathchar`-shell},
 \label{eq:f1onshell}
\end{align}
\begin{align}
 F_1 (\hat k^2) + F_2 (\hat k^2) &
 = \sum_{i,j,k = 1}^{3} \epsilon_{ijk}
 {\rm tr} \left[ (P_{\rm M})_i \Gamma_j \right]
\hat k_k  \bigg|_{\rm on\mathchar`-shell}.
\label{eq:f12onshell}
\end{align}

Now, by defining a complex variable $z = e^{i p_4}$, the integrals in
Eqs.~\eqref{eq:fE} and \eqref{eq:fM} can be written as contour
integrals on the unit circle. Inside the unit circle, the integrand
has poles of the external fermion propagators as well as a cut from
the fermion-photon intermediate states on the real axis of $z$.
The external fermion propagators form double poles,
\begin{align}
{1 \over s(p)^2 + m_f^2} {1 \over s(p+k)^2 + m_f^2}
 = {w(z) \over (z - z_*)^2 (z + z_*)^2},
\end{align}
where
\begin{align}
  z_* = - E + \sqrt{E^2 + 1},
\end{align}
is the location of the physical pole, and the pole at $-z_*$
represents that of the doubler degree of freedom. The function in the
numerator, $w(z)$, is a regular function at $z=\pm z_*$.
Each of the fermion propagators has the simple poles at the same
locations, $z = \pm z_*$, because of the relation, $p^2 = (p+k)^2$.
Since the matrices 
\begin{align}
    - i \gamma_4 (iE) -i {\bm \gamma} \cdot {\bm s} (p + k)  + m_f
\end{align}
and
\begin{align}
  - i \gamma_4 (iE) -i {\bm \gamma} \cdot {\bm s} (p)  + m_f
\end{align}
in the projection operators are residues of the external fermion
propagators, we see that the integrals in Eqs.~\eqref{eq:fE} and
\eqref{eq:fM} contain terms which are proportional to the desired form
factors in Eqs.~\eqref{eq:f1onshell} and \eqref{eq:f12onshell}. For
example, the first integral~\eqref{eq:fE} is evaluated as
\begin{align}
{\cal F}_E (t) & = {1 \over 2 \pi i} \oint_{|z|=1} dz \, {\rm tr} \left[
  \gamma_4 \hat G_4
  \right] z^{t - 1}
\nonumber                                                     \\
& = {d \over dz} \left( {\rm tr} \left[
  \gamma_4 \hat G_4
  \right]  (z - z_*)^2 z^{t - 1} \right) \bigg|_{z = z_*} + (z_* \to - z_*) + \cdots
\nonumber                                                     \\
& = t \left( {\rm tr} \left[
  \gamma_4 \hat G_4
  \right]  (z - z_*)^2 z^{t - 2} \right) \bigg|_{z = z_*} + (z_* \to - z_*) + \cdots,
\label{eq:fEintegral}
\end{align}
where $\cdots$ are terms which are subleading at large $t$. There are
contributions from the cut, but those are collections of simple poles
and thus there is no $t$ enhancement. Also, they are suppressed by
$e^{-m_\gamma t}$ at large $t$.
We see that the leading term is indeed the one which is proportional
to the right-hand-side of Eq.~\eqref{eq:f1onshell}.
The ratio in the numerator in Eq.~\eqref{eq:gfactor_formula} takes
care of most of the normalization factors except for the differences
of the coefficients in Eqs.~\eqref{eq:pE} and \eqref{eq:pM}, which are
taken care by the ratio in the denominator.
Therefore, by looking at the large $t$ behavior of
Eq.~\eqref{eq:gfactor_formula} and removing $O(1/t)$ suppressed
contributions, one can obtain the $g$ factor.

The Fourier transform of the three-point function is related to the
position-space quantity as
\begin{align}
  \int {d p_4 \over 2 \pi} G_\mu (p,k) e^{i p_4 t}
   & = {1 \over L^3} \langle 
   \psi ({\bf p}, t + t_2 ) \bar \psi (-{\bf p}-{\bf k}, t_2)
  \tilde A_\mu (k) \rangle e^{i k_4 t_2},
  \label{eq:form_factor_ft}
\end{align}
where
\begin{align}
  \psi ({\bf p}, t) 
  = \sum_{\bf x} \psi({\bf x}, t) e^{i {\bf p} \cdot {\bf x}}, \quad
  \bar \psi ({\bf p}, t) = \sum_{\bf x}
   \bar \psi({\bf x}, t) e^{i {\bf p} \cdot {\bf x}}.
\end{align}
The photon field $\tilde A_\mu$ is kept in the momentum space. The
right-hand-side of Eq.~\eqref{eq:form_factor_ft} is independent of
$t_2$ by the translational invariance. The phase factor $e^{i k_4
t_2}$ can be ignored for $k_4 = 0$. The variable $t$ simply represents
the separation of the fermion operators in the $x_4$ direction.

On the lattice, contributions from the doubler pole make the values of
${\cal F}_{E} (t)$ and ${\cal F}_{M} (t)$ vanishing for even integer
values of $t$, while, for odd integer $t$, the values are doubled
since the doubler contributions are the same as the physical ones. 
The values of ${\cal F}_{E}^{\rm (norm)} (t)$ and ${\cal F}_{M}^{\rm
(norm)} (t)$ are, on the other hand, vanishing for odd and even $t$,
respectively. This is problematic when we take the ratio for each
$t$ in Eq.~\eqref{eq:gfactor_formula}. In order to avoid the problem,
we replace the vanishing Fourier coefficients with the geometric means
of the next and previous coefficients before taking the ratio.
The factor of two caused by the doubler properly cancels in the
quantity~\eqref{eq:gfactor_formula}.

\section{Parameter choices}
\label{sec:parameters}

There are several requirements for the parameters in order to
approximate the continuum theory. It is summarized as
\begin{align}
 {1 \over T} \ll m_\gamma a \ll {|\hat {\bf k}|} a \ll m a \ll \Lambda_{\rm UV} a.
\end{align}
With the first condition $1/T \ll m_{\gamma} a$, we can get rid of the
fermion-photon intermediate states $\sim e^{-m_{\gamma} t}$ at large
$t$. The second condition $m_\gamma a \ll {|\hat {\bf k}|} a$ allows
us to regard that the photon is ``massless". The third condition
${|\hat {\bf k}|} a \ll m a$ corresponds to taking the external photon
momentum low enough compared to the fermion mass. The last condition
corresponds to setting  the UV cutoff sufficiently higher than the
typical scale of the observable.

The smallest momentum one can take on the lattice is
\begin{align}
 |\hat {\bf k}|_{\rm min} a = 2 \sin \frac{\pi}{L}.
 \label{eq:momchoice}
\end{align}
Let us fix $|\hat {\bf k}| = |\hat {\bf k}|_{\rm min}$ and
$\Lambda_{\rm UV} a = O(1)$.
A possible choice of $ma$ and $m_\gamma a$ would be such that
\begin{align}
  (ma)^2 = {|\hat {\bf  k}|^2 \over 4 m^2} 
  = {m_\gamma^2 \over {|\hat {\bf k}|^2}}.
  \label{eq:choice}
\end{align}
Each of them represents the size of errors from discretization, the
finite photon momentum, and the finite photon mass, respectively. We
also need to take a large enough $T$ such that $T \gtrsim (m_\gamma
a)^{-1}$.
The choice of Eq.~\eqref{eq:choice} leads to
\begin{align}
  (ma)^2 = \sin {\pi \over L}, \quad {\mbox{and}} \quad
  (m_\gamma a)^2 = 4 \sin^3 {\pi \over L}.
  \label{eq:Ldep}
\end{align}
Therefore, by fixing the relation in Eq.~\eqref{eq:choice}, one can
take $L \to \infty$ to approach the continuum theory while IR
regulators are sent to zero, {i.e.,} $\hat {\bf k} \to 0$ and
$m_\gamma \to 0$. The error of the $g$ factor at a finite $L$ is of
$O(\pi/L)$.

In fact, there are contributions which diverge as $m_{\gamma} \to 0$,
where the photon mass cannot be regarded as perturbations, at
intermediate steps. However, such contributions do not affect our
final results. These issues are discussed in Sec.~\ref{sec:9}.

\section{Numerical simulations}
\label{sec:numerical}

We perform a simulation by keeping up to the $e^{7}$ order, which
corresponds to the three-loop level in the Feynman diagrams. A set of
the Langevin equations are solved by the Runge-Kutta improved
algorithm~\cite{Helfand:1979} explained in
Appendix~\ref{sec:numerical_langevin}. We also employ the momentum
dependent step size to avoid the critical slowing down in the
evolution of the low-momentum modes~\cite{Batrouni:1985jn}.
The Langevin time $\tau$ is discretized with an interval $\epsilon$
times the momentum dependent factor,
\begin{align}
  {e^{-2 \hat k^2 / \Lambda_{\rm UV}^2} \over \hat k^2 + m_\gamma^2 },
\end{align}
for each momentum mode. The configurations are stored for each
$1/\epsilon$ steps of the Runge-Kutta algorithm.
We take $1/\epsilon = 50$ in this study. The accuracy of the
calculation is checked by comparing the $g$ factor and the $Z_3$
factor obtained by the stochastic method with those from the standard
Feynman-diagram method at one-loop level based on the same lattice
action in Eq.~\eqref{eq:action} and the same definitions in
Eqs.~\eqref{eq:gfactor_formula} and \eqref{eq:z3def}. 
We will show the comparison later.
The procedures of the Runge-Kutta improvement and the momentum dependent
step size are quite effective in reducing the number of steps
necessary for the convergence.

We use NEC SX-Aurora TSUBASA A500 at KEK for the generation of the
configurations and for the measurements of the correlation functions. 
Simulations are performed on the lattice with sizes 
$12^3 \times 24$,
$16^3 \times 32$,
$20^3 \times 40$, and
$24^3 \times 48$.
The parameters are set according to Eq.~\eqref{eq:choice}. The UV
cut-off is set to $(\Lambda_{\rm UV} a)^2 = 2.0$. We take the gauge
fixing parameter $\xi = 1$. We list the parameters and the number
of analyzed configurations in Tab.~\ref{tab:parameters}.

\renewcommand{\arraystretch}{1.06}
\begin{table}
  \begin{center}
    \begin{tabular}[t]{c|c|c|c|c|c|c}
       $L^3 \times T$ & $ma$ & $(\Lambda_{\rm UV}a)^2 $ & $\xi$ & $m_\gamma a$ &
      $\epsilon$ & $N_{\rm conf}$ \\ \hline
       $12^3 \times 24$ & $0.51$ & $2.0$ & $1.0$ & $0.26$ & $0.02$ & 4800 \\
       $16^3 \times 32$ & $0.44$ & $2.0$ & $1.0$ & $0.17$ & $0.02$ & 6400 \\
       $20^3 \times 40$ & $0.39$ & $2.0$ & $1.0$ & $0.12$ & $0.02$ & 7040 \\
       $24^3 \times 48$ & $0.36$ & $2.0$ & $1.0$ & $0.094$ & $0.02$ & 9600 \\
    \end{tabular}  
  \end{center}
  \caption{Simulation parameters. They are chosen according to the
  relation in Eq.~\eqref{eq:choice} so that the $L \to \infty$ limit
  is the continuum theory as in Eq.~\eqref{eq:Ldep}.}
  \label{tab:parameters}
\end{table}

We evaluate the two- and three-point functions by setting the external
photon momentum to be the minimal ones, ${\bf k} = (2\pi/L,0,0,0)$,
$(0,2\pi/L,0,0)$, and $(0,0,2\pi/L,0)$, and take an average of those
three measurements. The momentum of the external fermion is
accordingly set by ${\bf p} = -{\bf k}/2$, where half-integer wave
numbers are realized by imposing the anti-periodic boundary conditions
on the Dirac operators used in the measurements. The gauge
configurations are, on the other hand, generated by using the Dirac
operator with the periodic boundary conditions. The energy of the
external fermions, $p_4$, is scanned for all values, which are used
for the Fourier transform in Eqs.~\eqref{eq:fE} and \eqref{eq:fM}.
By the evaluation of Eq.~\eqref{eq:gfactor_formula}, we obtain the $g$
factor as a perturbative series of $e^2$. We reexpress the series by
that of $e_{\rm P}^2$ by reverting the relation in Eq.~\eqref{eq:eP}
as $e^2 = (Z_3^{(0)})^{-1} e_{\rm P}^2 - (Z_3^{(0)})^{-3} Z_3^{(2)}
e_{\rm P}^4 + \cdots$, where $Z_3=Z_3^{(0)}+Z_3^{(2)} e^2+\cdots$.
In that calculation, we use the $Z_3$ factor evaluated at $\hat k^2 =
\hat {\bf k}^2$, where $\hat {\bf k}$ is the external photon momentum
of the three-point function. The treatment is justified since we send
$\hat {\bf k} \to 0$ in the $L\to\infty$ limit.

\begin{figure}[t]
  \begin{center}
    \includegraphics[width=6.5cm]{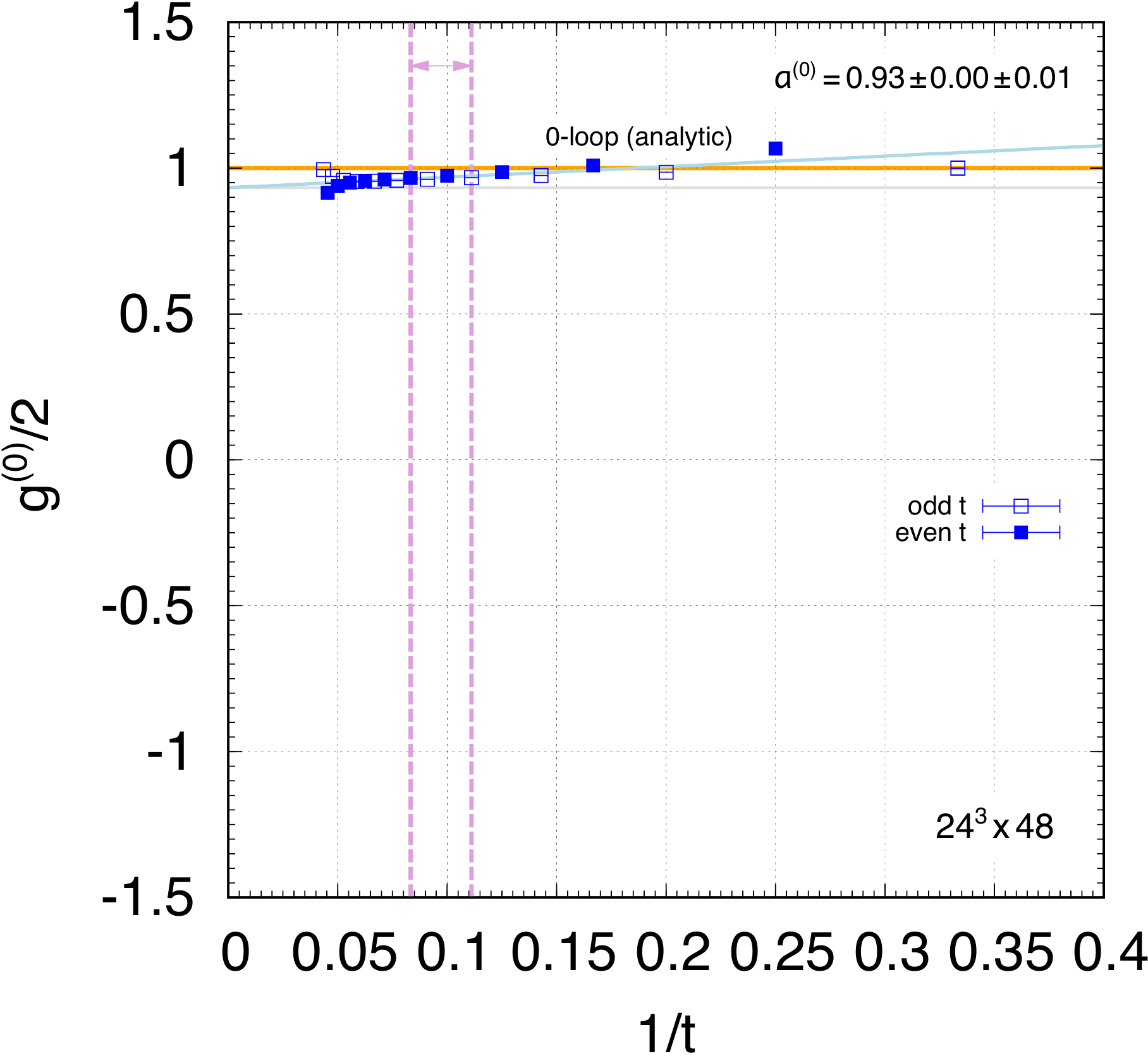}\hspace{10mm}
    \includegraphics[width=6.5cm]{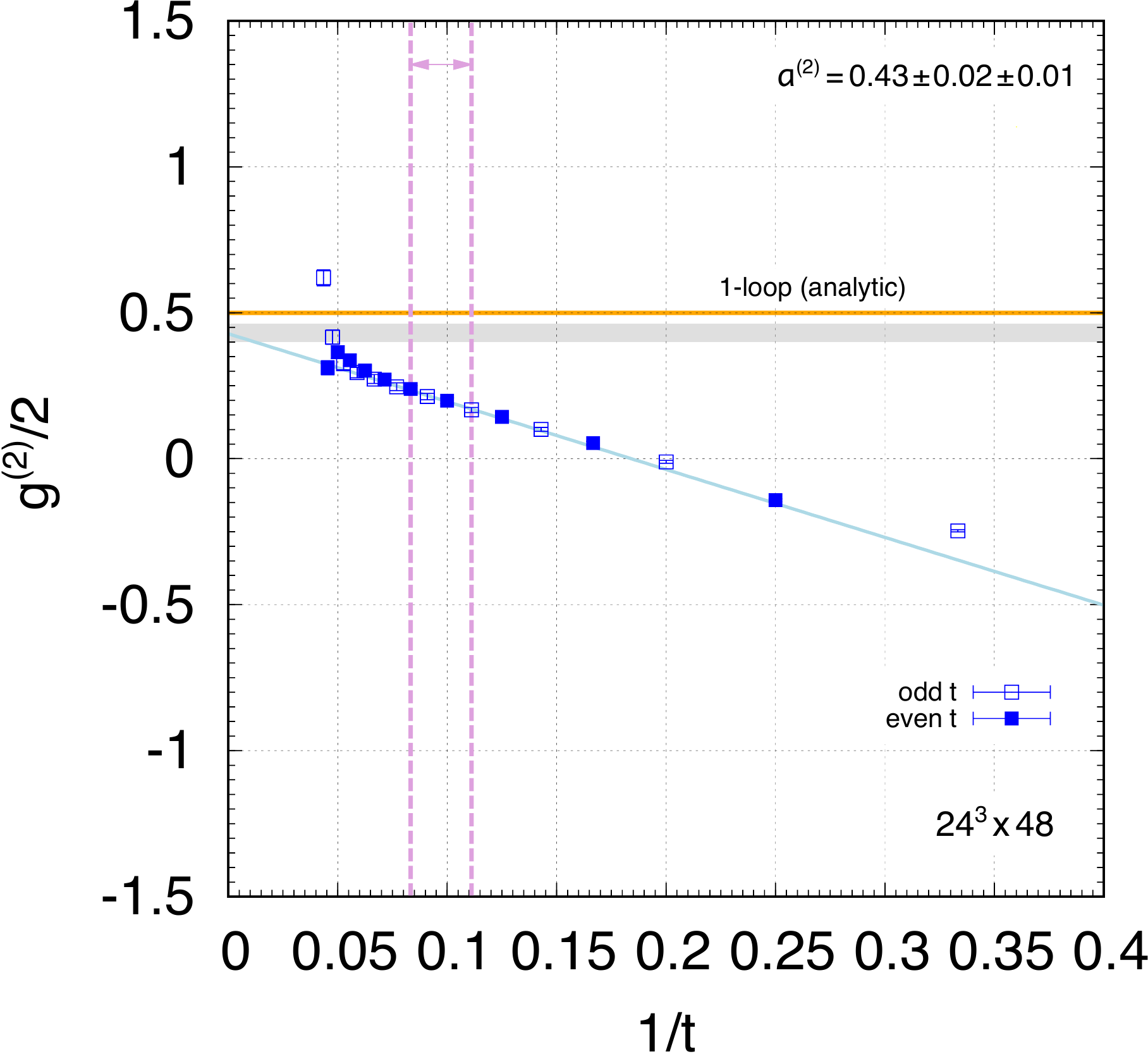}\\
    \vspace*{10mm}
    \includegraphics[width=6.5cm]{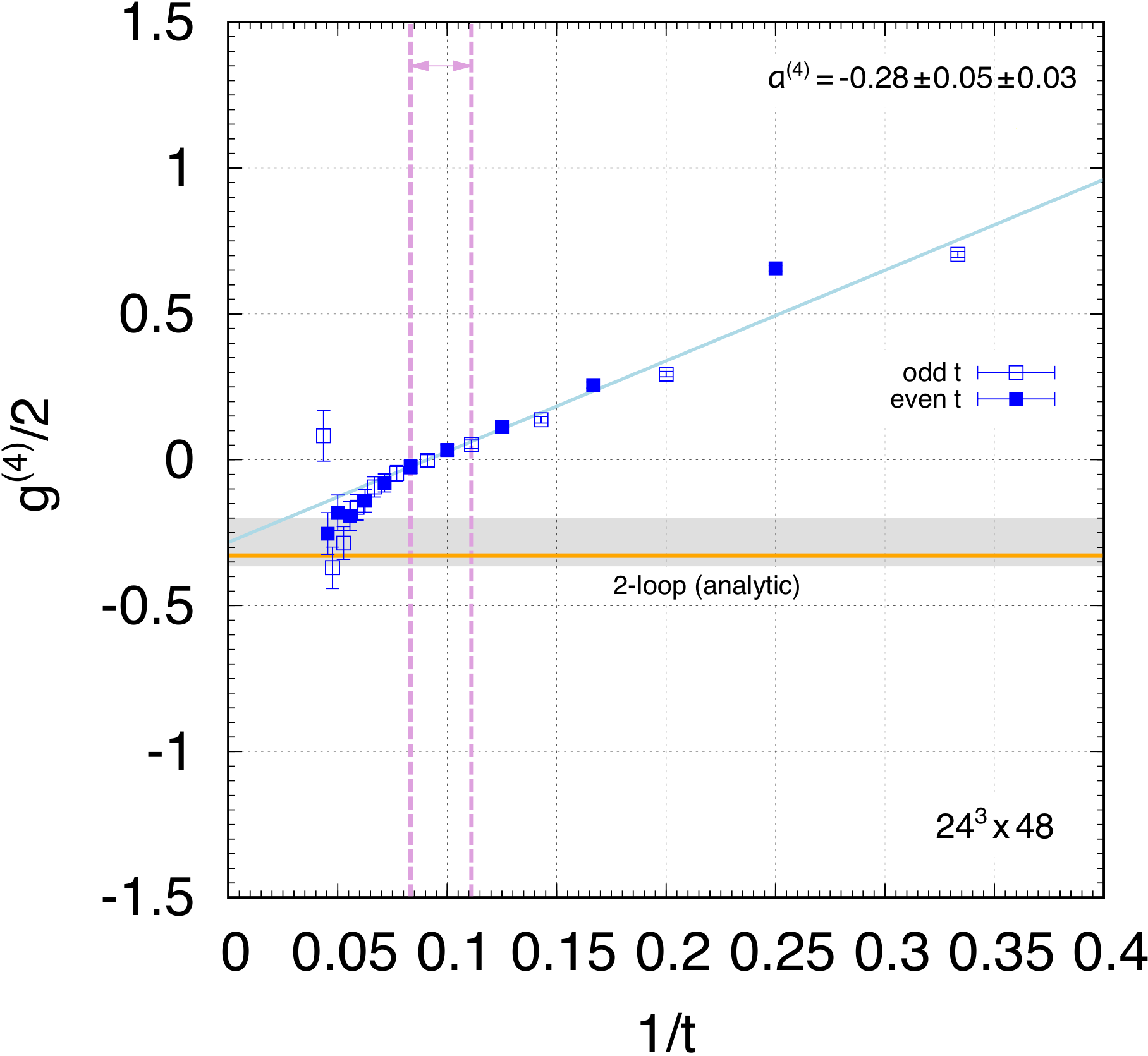}\hspace{10mm}
    \includegraphics[width=6.5cm]{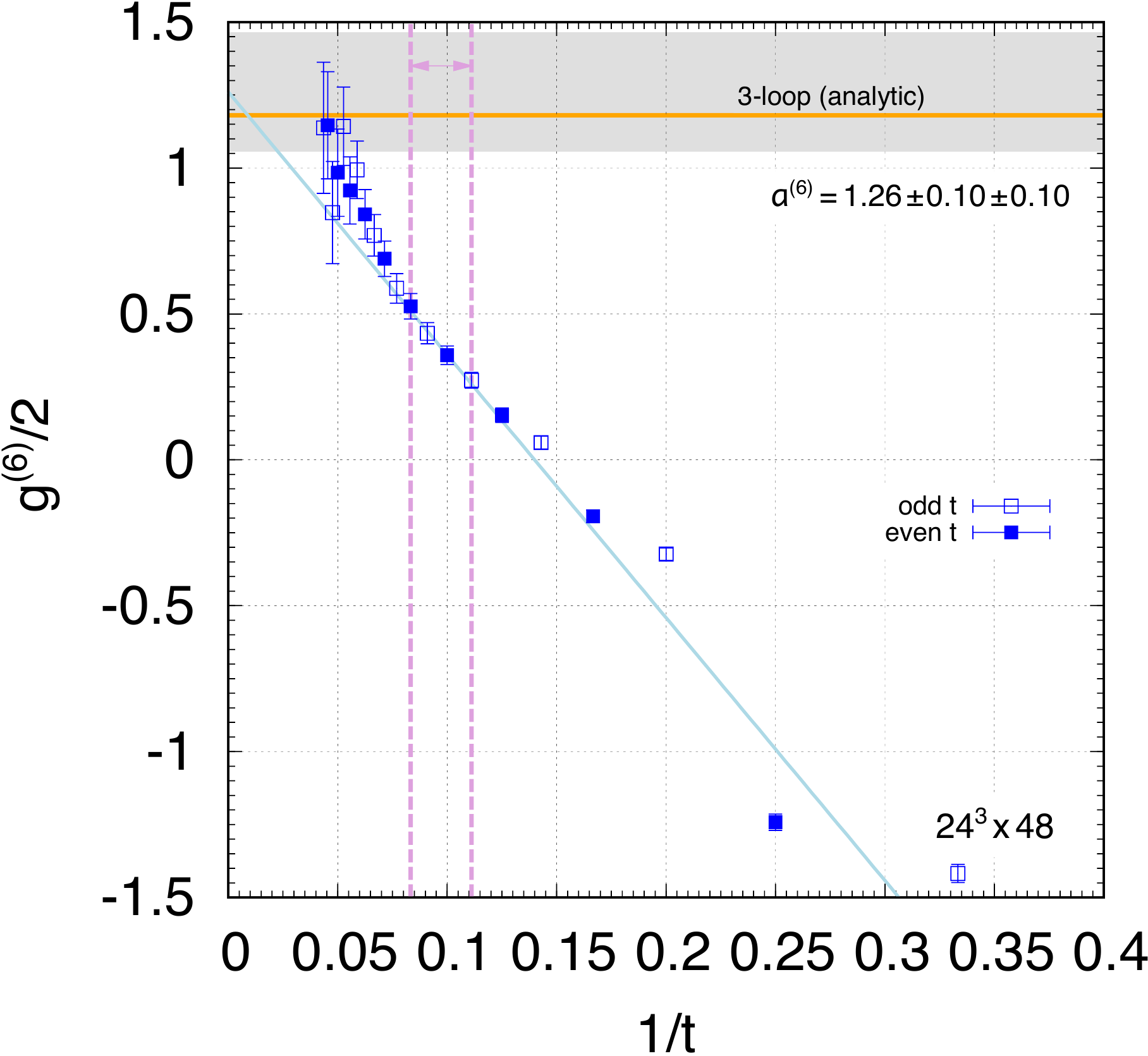}
  \end{center}

  \caption{Calculation of the $g$ factor on the $24^3 \times 48$
   lattice. Four points between the dashed lines are used for the
   extrapolations to $1/t \to 0$. Lines of the extrapolation are drawn
   by using the central values of $a^{(2n)}$ and $b^{(2n)}$ in
   Eq.~\eqref{eq:lines}.}
  \label{fig:24x48}
\end{figure}

We show in Fig.~\ref{fig:24x48} the function $g(t)$ in
Eq.~\eqref{eq:gfactor_formula} as a function of $1/t$ for the $24^3
\times 48$ lattice. The points are shown for $3 \leq t < L/2$.
At each order, coefficients, $g^{(2n)}(t) / 2$, of the expansion,
\begin{align}
  {g(t) \over 2} = 
    {g^{(0)}(t) \over 2}
  + {g^{(2)}(t) \over 2} \left( {\alpha \over \pi} \right)
  + {g^{(4)}(t) \over 2} \left( {\alpha \over \pi} \right)^2
  + {g^{(6)}(t) \over 2} \left( {\alpha \over \pi} \right)^3 + \cdots,
\end{align}
are plotted.
We see the expected behavior discussed in the previous section,
\begin{align}
  {g^{(2n)} (t) \over 2} \simeq a^{(2n)} + b^{(2n)} / t,
  \label{eq:lines}
\end{align}
where deviation can be seen at $1/t \lesssim 1/12 \sim 0.08$. We
extrapolate to the $t\to \infty$ limit, {\it i.e.}, $a^{(2n)}$, as
follows. We identify the linearly scaling region; the data at too
small $t$ are affected by subleading $t$ dependence, whereas
those at too large $t$ are contaminated by contributions
from backward propagations. We then select the points to be used in
the extrapolation to $1/t \to 0$ as $t = T/4-3$, $T/4-2$, $T/4-1$, and
$T/4$. We perform two linear extrapolations; one line connects the two
even points among these four points, and the other does the odd
points. We then take an average of the two extrapolated values. As we
discussed before, even and odd points are calculated in a different
manner due to vanishing Fourier coefficients by the doubler
contributions. We see in the figure that the even and odd points
separately behave as linear functions but eventually merge at large
$t$.
We take the difference between the even and odd extrapolations as a
systematic error.
By this procedure, one obtains the $g$ factor at a finite lattice
spacing. The lines $a^{(2n)} + b^{(2n)} / t$ obtained in this way are
shown in the figure. The grey shaded regions are the values of
$a^{(2n)}$ with statistical and systematic errors whereas the
horizontal lines show the known coefficients by the standard Feynman
diagram method:
\begin{align}
  a^{(0)} = 1, \quad
  a^{(2)} = 0.5, \quad
  a^{(4)} = -0.328\dots \quad
  a^{(6)} = 1.18\dots
  \label{eq:continuum}
\end{align}
We see that our parameter choice at $L=24$ is not too far from the
continuum limit.

Repeating the same calculations for other sizes of lattices, one can
see the $L$ dependence of the $g$ factors. The results are listed in
Table~\ref{tab:results}.
The statistical errors are estimated by the jackknife method with the
bin size $n_{\rm bin}=10$.
With the parameter relations in Eq.~\eqref{eq:choice}, it is expected
for large $L$ that the values approach to the continuum limit as 
\begin{align}
  g(L) \sim  g(\infty) + {c \pi \over L}, 
  \label{eq:error}
\end{align}
with an $O(1)$ coefficient, $c$.
The coefficient, $c$, however, can have logarithmic dependence on $L$
reflected by the logarithmic UV divergences in the relation between
$m_f$ and $m$. Since the continuum limit $a\to0$ should be taken with
the physical fermion mass $m_f$ fixed, the actual discretization error
is $O(m_f^2 a^2)$ rather than $O(m^2 a^2) \sim O(\pi / L)$. The size
of the logarithmic correction is controlled by the UV cut-off
parameter $\Lambda_{\rm UV}$ in the action. We take $(\Lambda_{\rm UV}
a)^2 = 2.0$ which makes the logarithmic correction to be of $O(1)$.
For $\Lambda_{\rm UV} a \to \infty$, the continuum limit gets far away
as we discuss later.

\begin{table}
  \begin{center}
    \begin{tabular}[t]{l|l|l|l|l}
      $L^3 \times T$   & $a^{(0)}$ & $a^{(2)}$ & $a^{(4)}$  & $a^{(6)}$  \\ \hline
      $12^3 \times 24$ & $0.85(0)(4)$ & $0.46(2)(6)$ & $-0.54(2)(11)$ & $1.14(4)(17)$  \\
      $16^3 \times 32$ & $0.90(0)(2)$ & $0.45(2)(2)$ & $-0.43(3)(6)$ & $0.96(6)(6)$  \\
      $20^3 \times 40$ & $0.92(0)(1)$ & $0.43(2)(2)$ & $-0.38(4)(5)$ & $1.03(8)(8)$  \\
      $24^3 \times 48$ & $0.93(0)(1)$ & $0.43(2)(1)$ & $-0.28(5)(3)$ & $1.26(10)(10)$ \\
      \hline
      (analytic) & 1  & 0.5 & $-0.328\dots$ & $1.18\dots$ \\
    \end{tabular}
  \end{center}
  \caption{Simulation results. Parameters are listed in
  Table~\ref{tab:parameters}. The first and second parentheses
  represent statistical and systematic errors, respectively. We
  estimate the total errors as the sum of two numbers. The values in
  the continuum theory are listed in the last row.}
  \label{tab:results}
\end{table}

\begin{figure}[]
  \begin{center}
    \includegraphics[width=6.5cm]{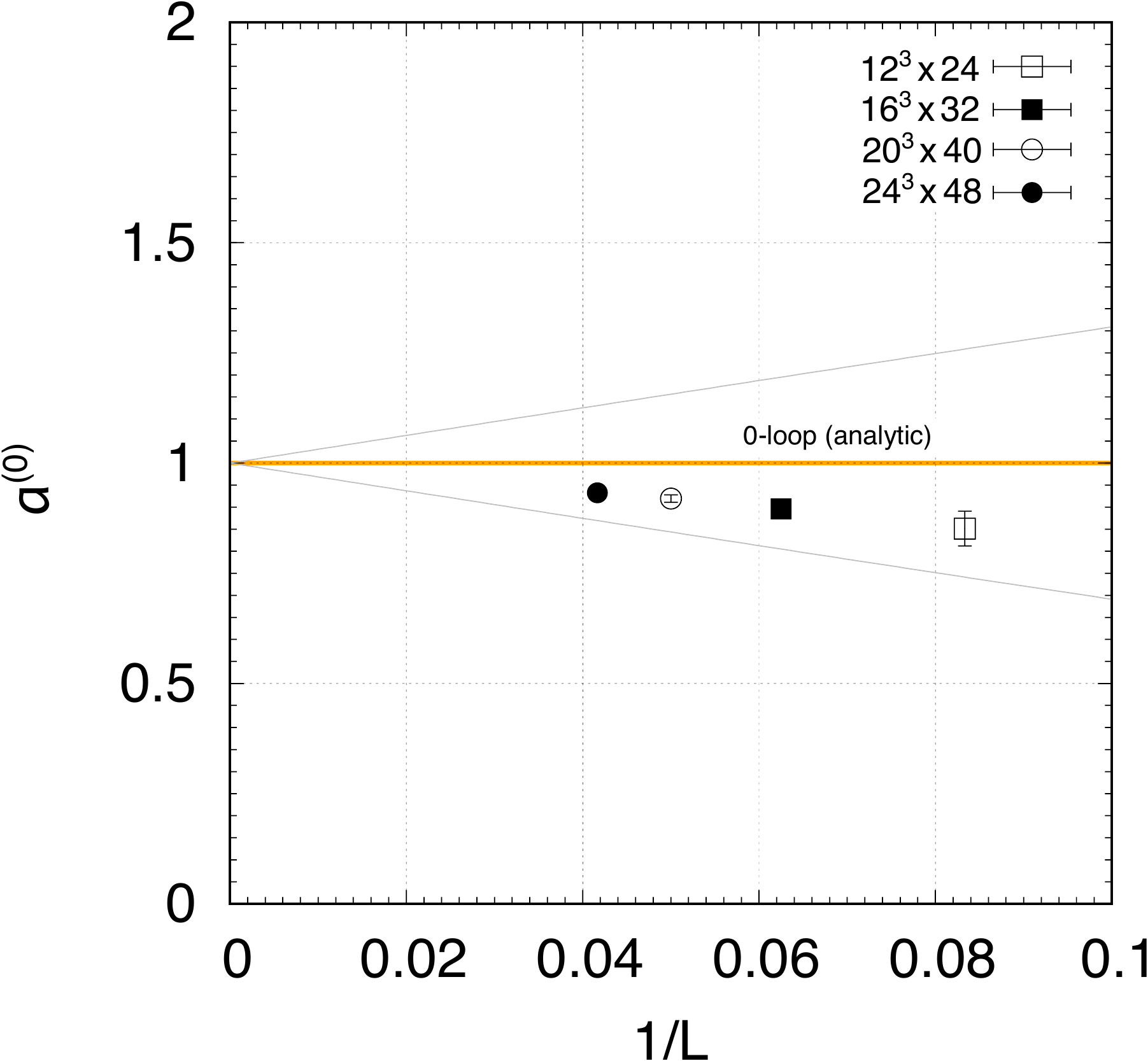}\hspace{10mm}
    \includegraphics[width=6.5cm]{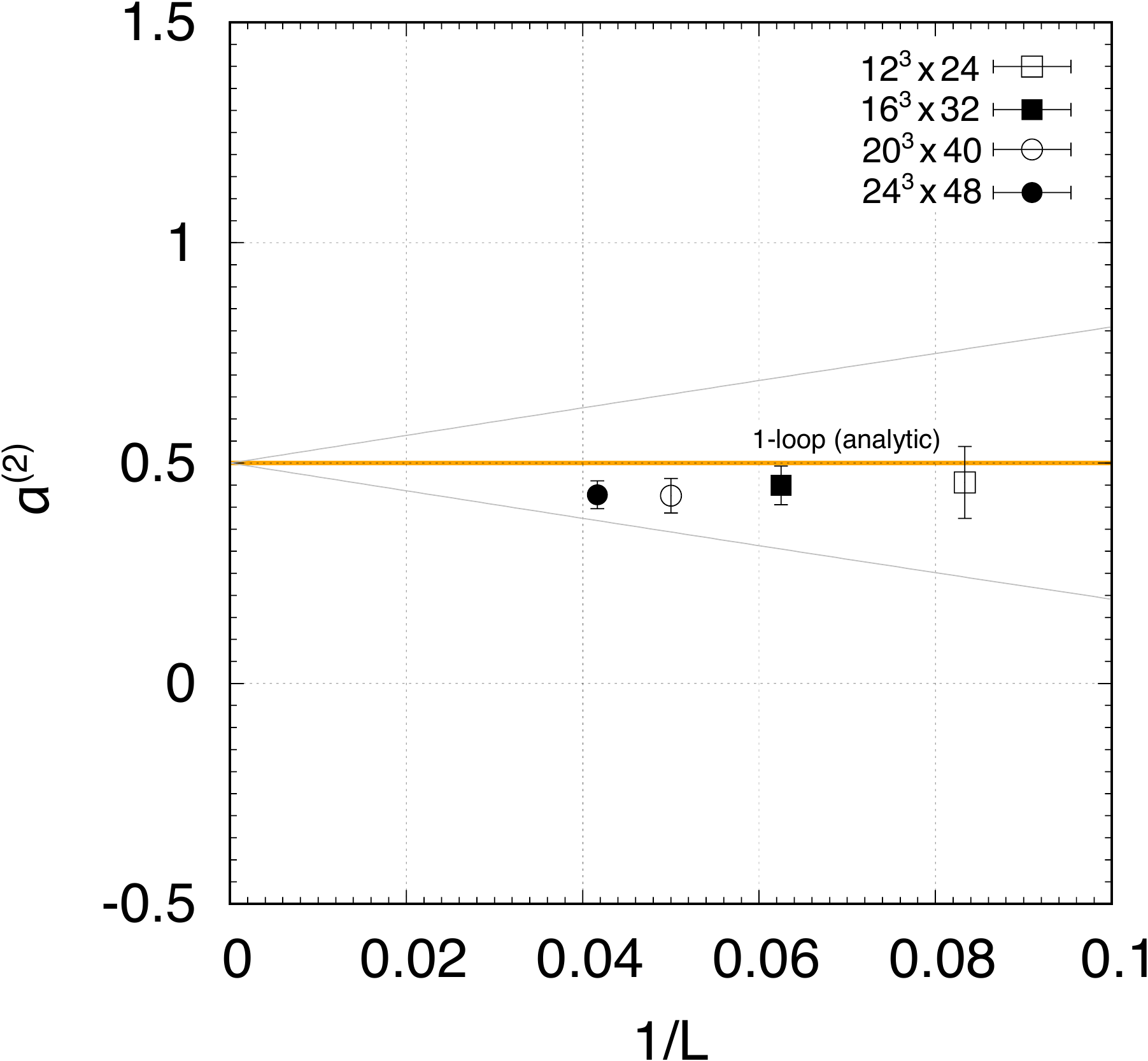}\\
    \vspace*{10mm}
    \includegraphics[width=6.5cm]{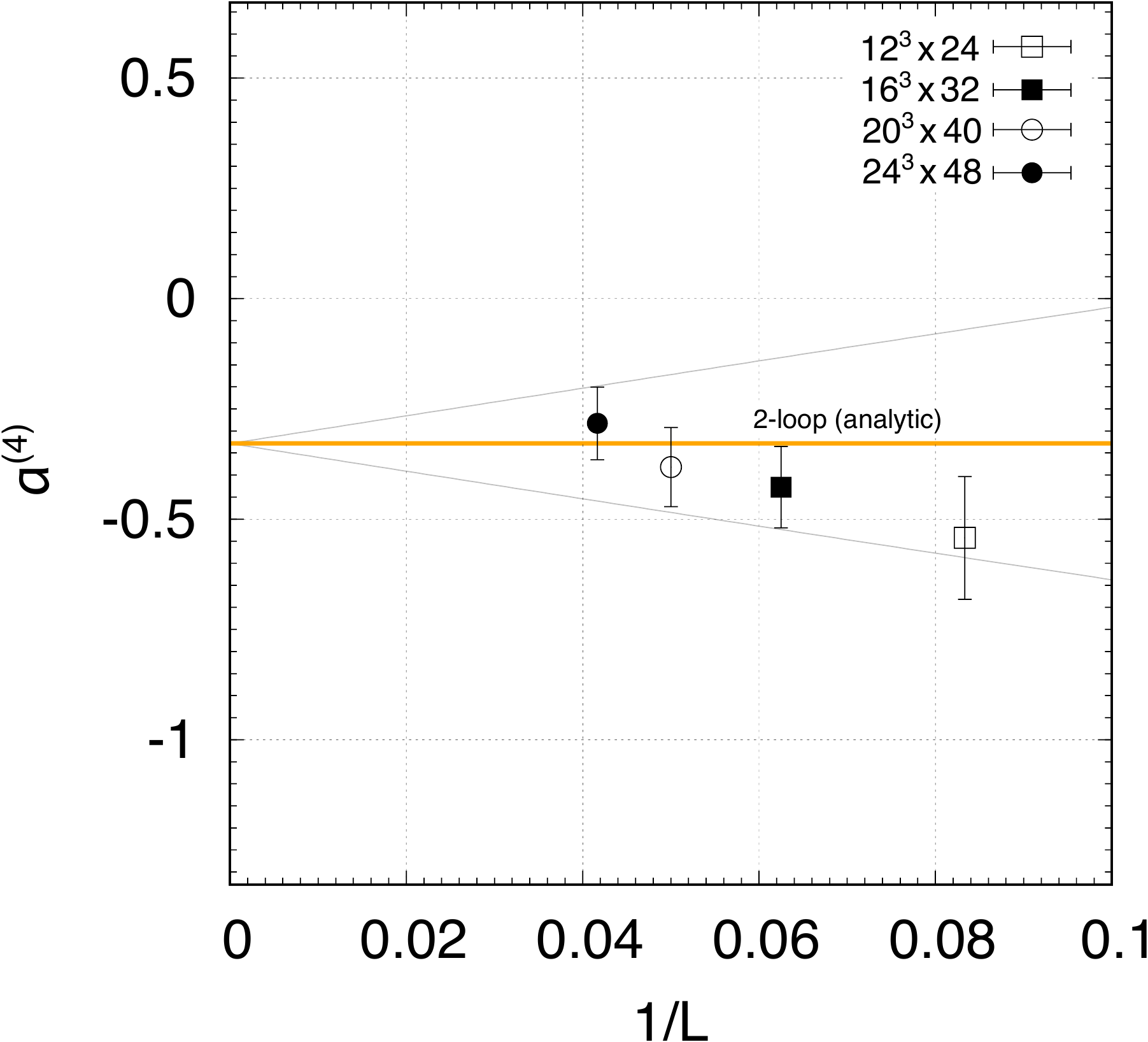}\hspace{10mm}
    \includegraphics[width=6.5cm]{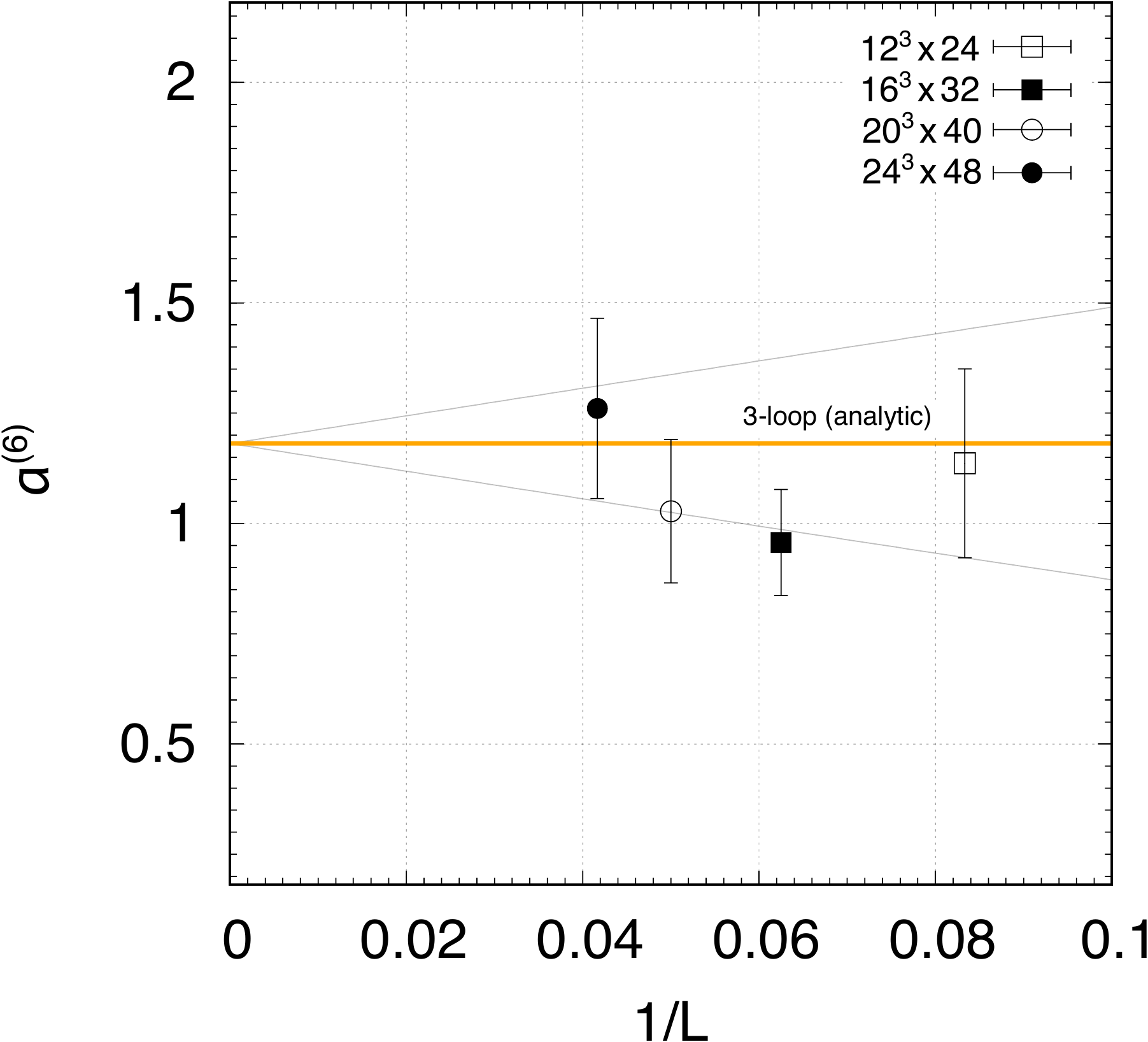}
  \end{center}

  \caption{$L$ dependence of the perturbative coefficienets of the $g$
    factor on the lattice. $1/L \to 0$ corresponds to the continuum
    limit by fixing the parameter relations in Eq.~\eqref{eq:choice}.
    The grey lines represent expected sizes of the discretization
    errors, $\pm (ma)^2$.}
  \label{fig:Ldep}
\end{figure}

We show in Fig.~\ref{fig:Ldep} the $L$ dependence of the $g$ factor
coefficients, $a^{(2n)}$, for each order of perturbations. The
horizontal lines are again the known analytic results in the continuum
theory.
The grey lines represent the typical sizes of the discretization
errors, $\pm (ma)^2$.
We see a good agreement with the analytic results within the expected
uncertainties. At the tree level, there is no statistical error even
though we use $A_\mu^{(0)}$ in Eq.~\eqref{eq:gmu}, which fluctuates in
the Langevin evolutions. This is because at the tree-level
$A_{\mu}$ appears only at the external line 
and its fluctuation is exactly canceled when the photon propagator is removed 
as in Eq.~\eqref{eq:photon_removal} in each configuration.
Also, in the renormalization of the charge in Eq.~\eqref{eq:eP}, we
use the analytic expression, $Z_3^{(0)} (\hat k^2) = e^{-2 \hat k^2 /
\Lambda_{\rm UV}^2}$, for the tree-level value.
The deviation from the continuum theory seen in the figure is due to
the $O(a^2)$ correction in the fermion-fermion-photon vertex in the
lattice Dirac operator.
We do not try to obtain the extrapolated values in the $1/L \to 0$
limit as a reliable estimation seems to require points with larger
lattice sizes to control the logarithmic correction of the form $(1/L)
\ln L$.

\begin{figure}
\begin{center}
\includegraphics[width=11cm]{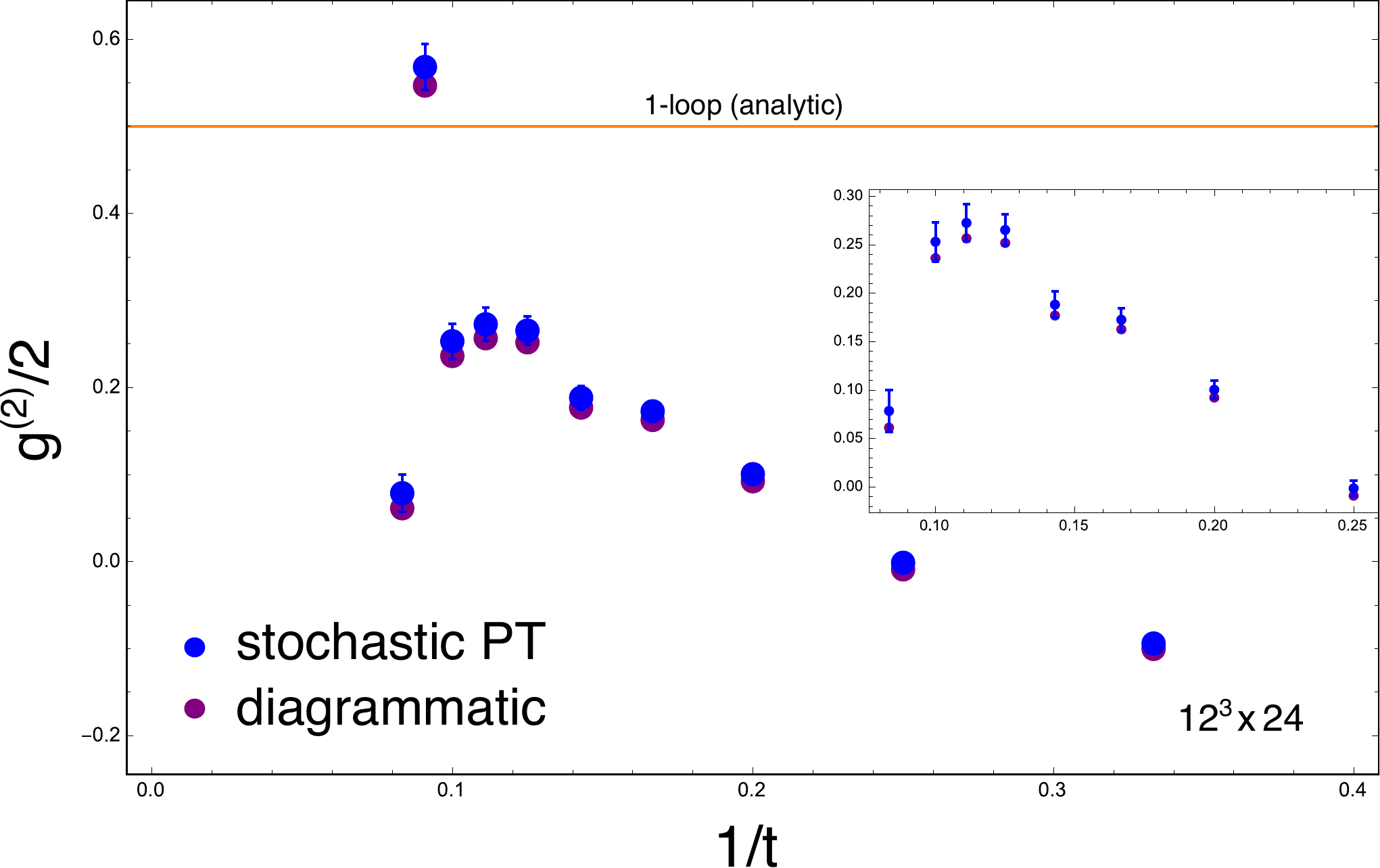}
\end{center}
\caption{Comparison of the function $g(t)$ at one-loop level between
the stochastic perturbation theory (blue) and the conventional lattice
perturbation theory using Feynman diagrams (purple) on $12^3 \times
24$ lattice. Magnified figure is also shown.}
\label{fig:StochasticFeynman}
\end{figure}

To confirm the validity of our calculation, e.g., the validity of the
discretized Langevin steps, we also perform one-loop calculations with
the method using Feynman diagrams. First, we compare the wave-function
renormalization factor of the photon, $Z_3$. We obtain
$Z_3^{-1}=1.30728 + 0.563(3) (e^2/(4\pi^2)) + \cdots$ with the
stochastic method, whereas the diagrammatic calculation gives
$Z_3^{-1}=1.30728+0.555891 (e^2/(4\pi^2))+\cdots$. 
While discrepancy beyond the statistical error is seen in the one-loop
coefficient, agreement at the level of a percent is confirmed.
The vacuum polarization is obtained by the expansion of the fermion
determinant in the stochastic calculation. One may need to increase
the number of Langevin steps and/or to generate more noise spinors in
Eq.~\eqref{eq:noise} if better precision is necessary.
The function $g(t)$ in the $12^3 \times 24$ lattice with a
diagrammatic method is also compared with the stochastic result in
Fig.~\ref{fig:StochasticFeynman} at the one-loop level. We find good
consistency within the statistical uncertainties.

\begin{figure}[t]
  \begin{center}
    \includegraphics[width=6.5cm]{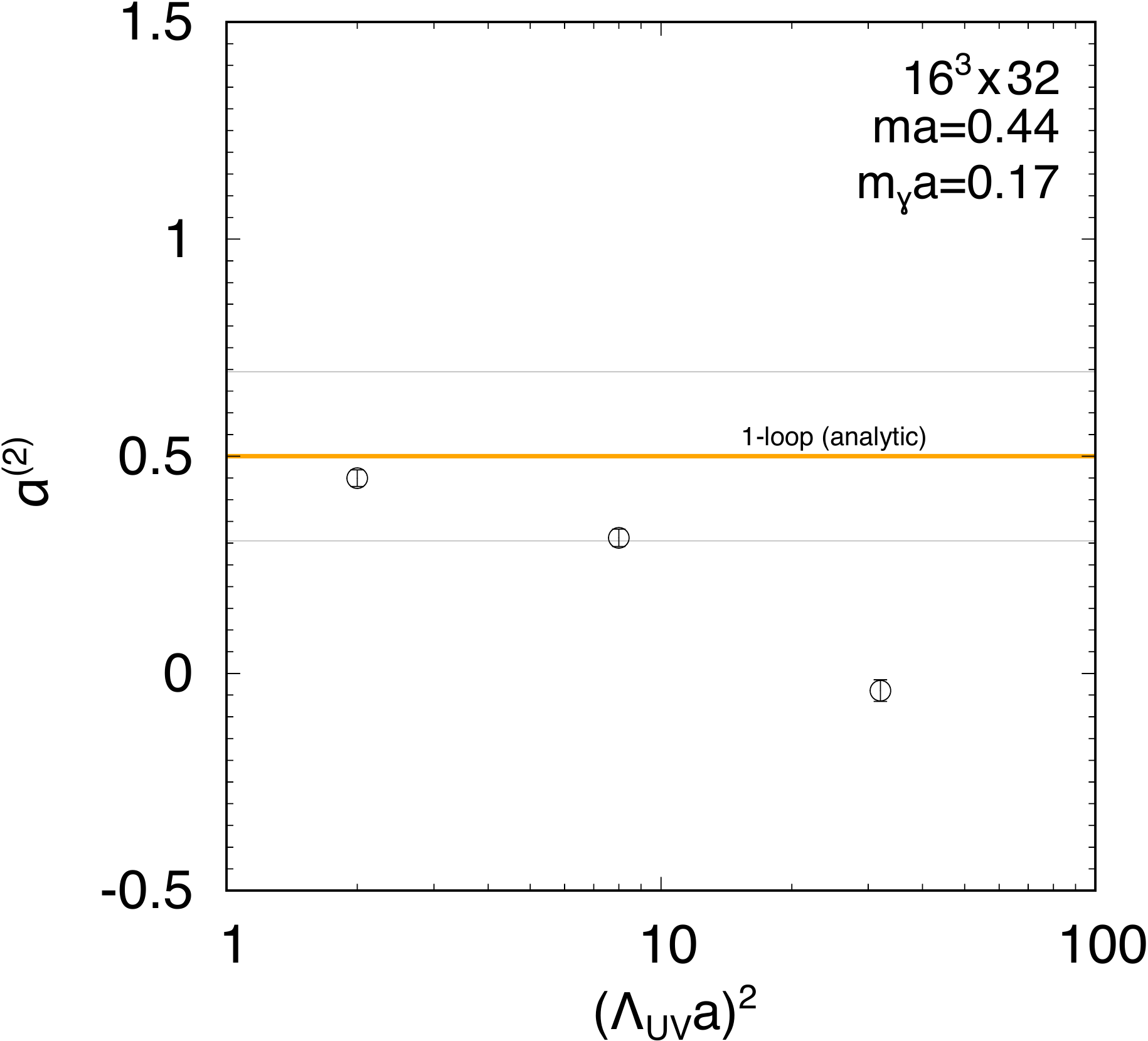}
  \end{center}

  \caption{$\Lambda_{\rm UV}$ dependence of the one-loop coefficient,
  $a^{(2)}$, on the $16^3 \times 32$ lattice. The grey lines represent
  the typical size of discretization errors, $\pm (ma)^2$, when we
  ignore the logarithmic correction. Only the statistical errors are
  shown.}
  \label{fig:UVdep}
\end{figure}

The $a^{(2n)}$ factors depend on the UV cutoff parameter $\Lambda_{\rm
UV}$ at finite lattice spacings.
We show in Fig.~\ref{fig:UVdep} the $\Lambda_{\rm UV}$ dependence of
the one-loop coefficient, $a^{(2)}$, on the $16^3 \times 32$ lattice.
We take $(\Lambda_{\rm UV} a)^2$ to be $2.0$, $8.0$, and $32.0$,
respectively, while keeping other parameters to be the same as the
ones listed in Table~\ref{tab:parameters}. The same numbers of
configurations (6,400~confs) are analyzed for each $({\Lambda_{\rm
UV}} a)^2$.
The grey lines represent the typical discretization errors, $\pm
(ma)^2$, as before.
In principle, one can take any value of $\Lambda_{\rm UV} a \neq 0$ as
long as we take the continuum limit in the end, but too large
$\Lambda_{\rm UV}a$ makes the continuum limit far away as we can see
from the figure.
This demonstrates that the introduction of a UV regulator is
essential in this calculation, otherwise we would need much larger
lattice volumes to see the convergence to the continuum values.

One may also employ the gradient flow~\cite{Luscher:2010iy} to reduce
the UV fluctuations rather than modifying the action. Also, one can
suppress the UV modes by modifying the noise term in the Langevin
equation to be momentum dependent as discussed in
Ref.~\cite{Bern:1986dj}. Those two treatments are equivalent to the
method of the modified action at the tree level, and may be useful in
QCD or in general theories where a gauge invariant UV regularization
cannot be simply implemented.
An advantage of the modified action is that one can manifestly
maintain the equivalence with the path integral formalism, and thus
the way to going back to the original theory is theoretically clear.

\section{IR and finite volume effects}
\label{sec:9}

As we modify the QED action to include the finite photon mass, the
gauge invariance is broken. In particular, the effects of the photon
zero mode should be carefully treated. We also have finite volume
effects in the calculation of the function $g(t)$. We discuss those
issues in this section.

We first discuss the effects of the photon zero mode. Although we
assumed that $\hat{G}_{\mu}$ and $\hat{G}^{\rm (norm)}_{\mu}$ in
Eqs.~\eqref{eq:photon_removal} and \eqref{eq:Gnorm} have the double
poles due to the on-shell divergences of the external fermions, they
actually have higher poles. These poles show up, in the language of
Feynman diagrams, as an IR divergence where internal fermion
propagators hit on-shell when the photon loop momentum vanishes.
For instance, at one loop, since we have four fermion propagators at
most, $\hat{G}_{\mu}$ has a quartic pole.
With this higher pole, the functions $\mathcal{F}_E(t)$ and
$\mathcal{F}_M(t)$ behave as $\sim t^3 z_*^{t-2}$ with a prefactor of
$O(1/(L^3 T m_{\gamma}^2))$.
This unwanted growth of the functions are worse for smaller
$m_{\gamma}$. Moreover, since we fix $m_\gamma^2 \sim 4(\pi / L)^3$ in
Eq.~\eqref{eq:Ldep}, the effects do not disappear in the $L \to
\infty$ limit at large $t$.

However, these seemingly problematic contributions are cancelled in
the function $g(t)$ of Eq.~\eqref{eq:gfactor_formula}. This can be
understood from the fact that the contributions from the photon zero
mode is factorized as an overall factor $\propto e^{-xt^2}$ with
$x=e^2/(2 L^3 T m_\gamma^2)$~\cite{Endres:2015gda, Endres:2015vpi} in
the fermion two-point functions and the three-point function we
consider.

The effects of the backward propagation in the correlation functions
are more serious in the extraction of the $g$ factor we developed. 
For instance in Eq.~\eqref{eq:fE}, related to the symmetry of the loop
{\it{sum}} under $t \to T-t$, the leading $t$ dependent part should be
accompanied with the backward propagating contribution:
\begin{align}
\mathcal{F}_E(t) 
&\sim t z_*^{t} e^{-xt^2} +(T-t) z_*^{(T-t)} e^{-x(T-t)^2} \nonumber \\
&\sim t e^{-Et} e^{-xt^2} \left[ 1+\left(\frac{T}{t}-1 \right) 
e^{-E (T- 2t)} e^{-xT(T - 2t)}\right] \nonumber \\
&=t e^{-Et}  e^{-xt^2} \left[1+\left(\frac{T}{t}-1 \right) 
e^{- E_0 (T-2t)} \left(1- e^2
\left(E_1 + {1 \over 2 L^3 m_\gamma^2} \right)
  (T- 2t) +\cdots
  \right) \right].
  \label{eq:fEbehavior}
\end{align}
By ignoring the backward propagation, one obtains $\mathcal{F}_E(t)
\sim t e^{-Et}$ up to the $e^{-xt^2}$ factor which is to be canceled
in the ratio.
The extra terms disturb this behavior.
In principle, one can fit the function $g(t)$ by using the above
expected behavior. Rather than trying such an analysis, for
simplicity, we choose a strategy to look for a region of the
extrapolation where the extra term is not important.
In future works, it is worth trying such an analysis.

The extra contributions are more and more important for higher orders.
In the last line, we express the correction term as a perturbative
series.
The fermion energy as well as the factor of the photon zero mode are
expanded as $E=E_0+e^2 E_1+\cdots$ and $e^{-xT(T-2t)} = 1 -
(e^2/(2L^3m_\gamma^2))(T-2t) + \cdots$, respectively.
One can see that the powers of $(T-2t)$ grow as $(T-2t)^{n}$ at the
$e^{2n}$ order.
Since there is an overall suppression factor $e^{-E_0 (T-2t)}$, one
can avoid this problem for sufficiently large $T$ while keeping the
region of extrapolation to be $t \ll T/2$. Nevertheless, for a fixed
$T$, there is a certain perturbative order beyond which the
extrapolation by Eq.~\eqref{eq:lines} is not reliable.

\begin{figure}[bhpt]
  \begin{center}
    \includegraphics[width=6.5cm]{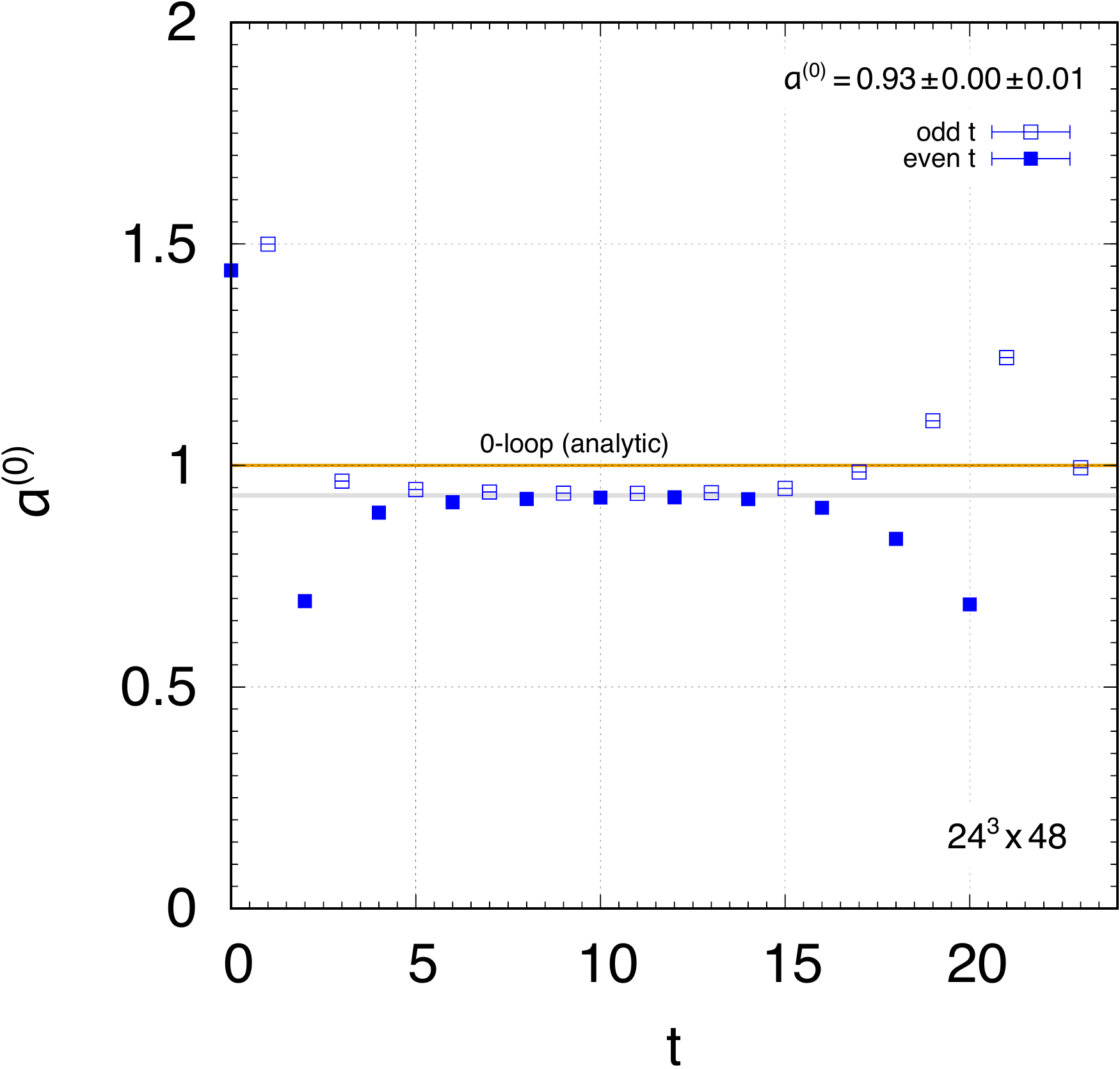}\hspace{10mm}
    \includegraphics[width=6.5cm]{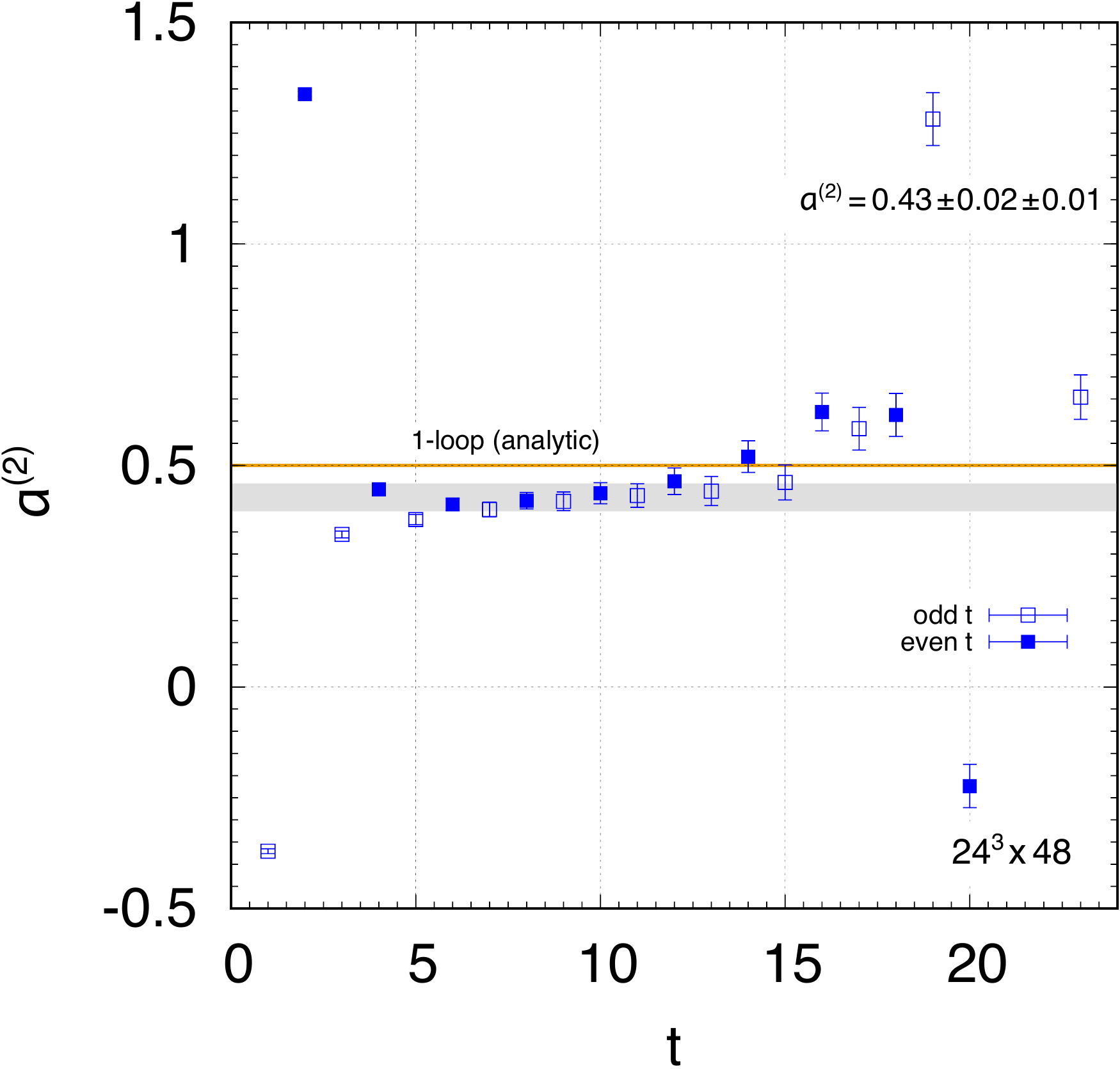}\\
    \vspace*{10mm}
    \includegraphics[width=6.5cm]{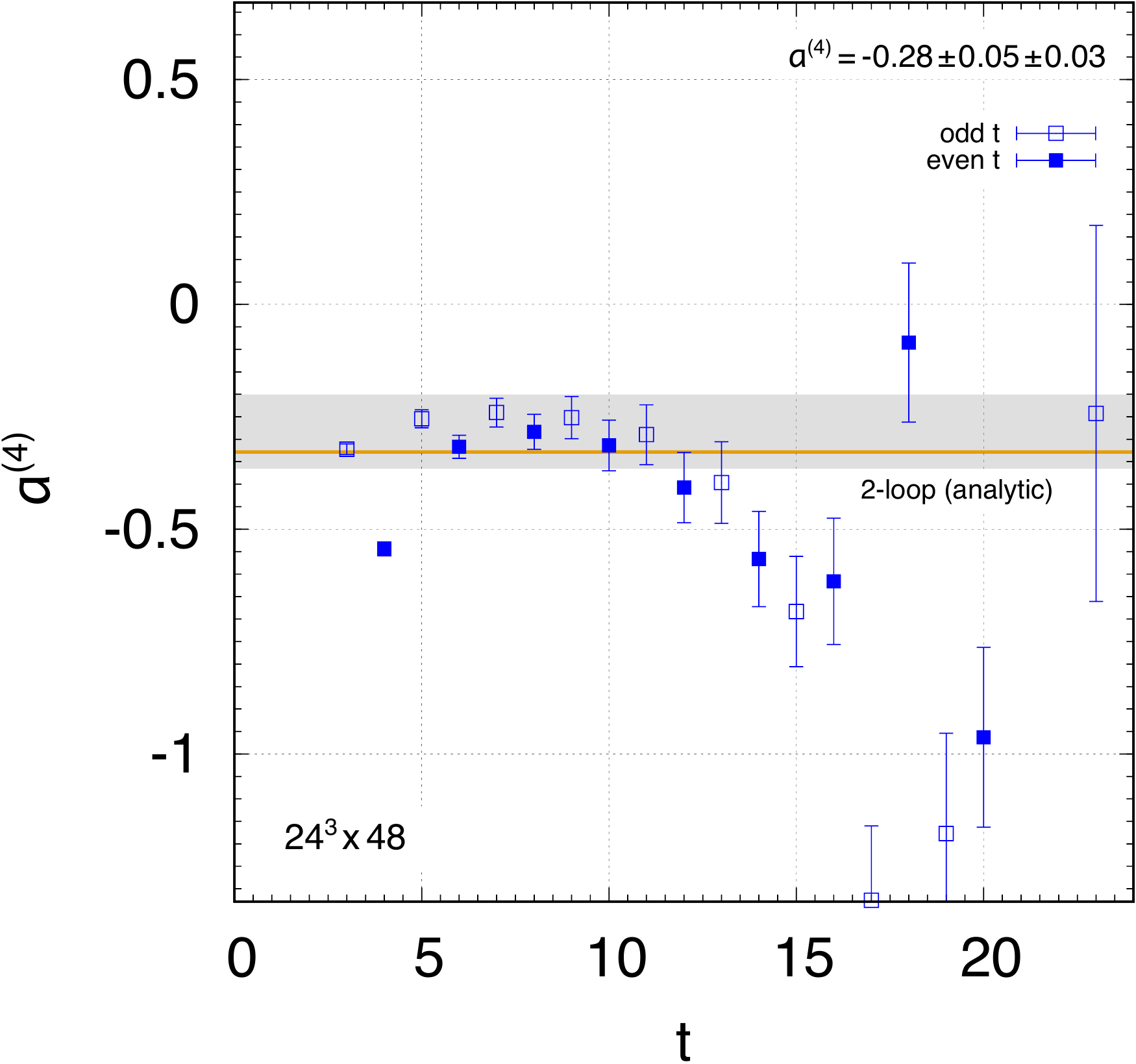}\hspace{10mm}
    \includegraphics[width=6.5cm]{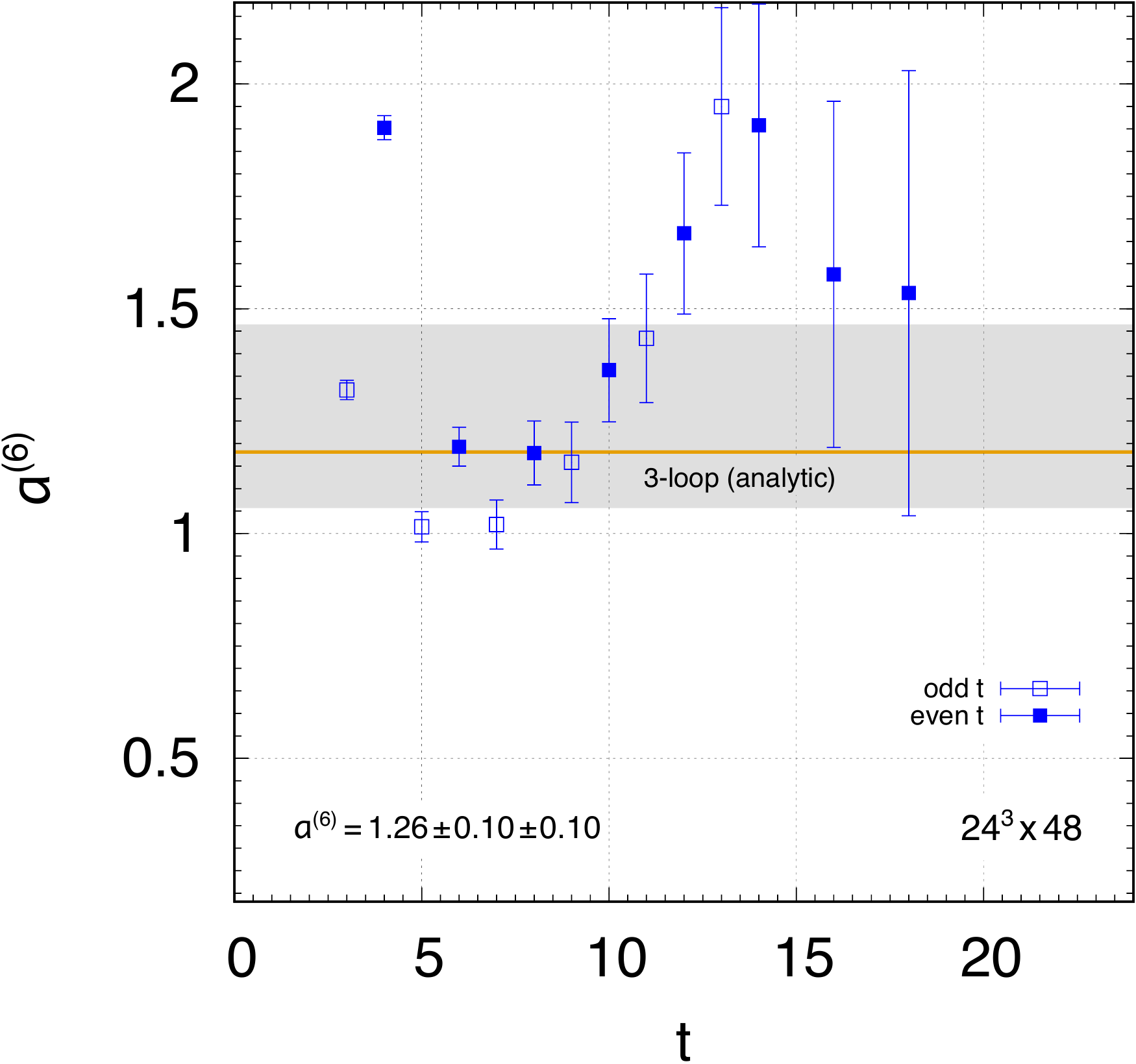}
  \end{center}
\caption{Dependence of the $n$-loop perturbative coefficients
$a^{(2n)}$ on the points used in the extrapolation of $g(t)$. The
$a^{(2n)}$ factors obtained by the extrapolations based on
Eq.~\eqref{eq:lines} by using $g(t)$ at $t$ and $t+2$ are shown. The
grey bands correspond to our results in Table~\ref{tab:results}.}
  \label{fig:fitrangedep}
\end{figure}

Such a limitation is already observed in Fig.~\ref{fig:24x48}, where
the deviation from the straight line starts to show up at smaller $t$
for higher orders.
In order to visualize the deviation, we show in
Fig.~\ref{fig:fitrangedep} the $a^{(2n)}$ factors obtained by the
extrapolation of points at $t$ and $t+2$.
Up to the two-loop level, we see clear plateaus, which indicate that
the results are not significantly sensitive to the range of
extrapolations. The plateaus are indeed consistent with our results
with the estimated error given in Table~\ref{tab:results}. 
At the three-loop order, the plateau seems to start disappearing. A
simulation with larger $T$ will be necessary to perform a more
reliable estimate, especially for computation beyond three loops.

\section{Summary}
\label{sec:summary}

We apply the numerical stochastic perturbation theory to QED and
evaluate the perturbative coefficients of the $g$ factor of the
electron up to $O(\alpha^3)$. We demonstrate that the method
reproduces the known results from the diagrammatic method, and expect
that it may be able to confirm or predict the higher order
perturbative coefficients. Since the method does not rely on the
Feynman diagrams, going to higher orders is a matter of computational
time and the memory footprint. For example, the increase of the
computational cost to accumulate the same number of configurations at
the five-loop level is about a factor of two compared to the present
work. Although we need more statistics as well as more data points at
larger lattice volumes to approach the continuum limit with controlled
systematic effects, the results obtained on a small-scale computer in
this work are quite encouraging.

The numerical stochastic perturbation theory has been successfully
applied to QCD to compute the quantities such as the vacuum energy
density and the heavy-quark self-energy, which can be naturally
defined on the Euclidean space-time.  The challenge we faced in this
work is to obtain the physical quantity, $g-2$, defined for an
electron on its mass-shell. Thus, an analytic continuation from the
Euclidean momentum to the on-shell electron is inevitable. We
introduce the Cauchy integral of the vertex function in terms of the
imaginary temporal momentum $p_4$ and design an appropriate
combination, from which the $g$-factor can be obtained at large
Euclidean time separations. 

The main practical limitation of the method is to separate various
energy scales including the electron mass, the external photon
momentum and the photon mass. The latter two of these have to be sent
to zero in order to obtain the physical result. The photon mass is
introduced to avoid an infrared divergence due to finite volume, and
it also plays the role to lift the energy of multi-particle states
made of an electron and photon so that the single electron state can
be easily isolated. Therefore, the extrapolation to the massless
photon limit has to be done with great care. In this work, we set the
parameters such that these energy scales are separated equally in the
logarithmic scale. The systematic error then scales rather slowly,
$\sim 1/L$, as a function of the lattice extent $L$. The largest
lattice volume in this work is $L=24 $; we expect that the simulation
at $L\sim 128$ would be possible on the large-scale machines currently
used for modern QCD simulations.

For the complete evaluation of the $g-2$ of the electron or the muon,
we need to include fermions with different masses to incorporate the
mass-dependent contributions. It is straightforward to include those
in the Lengiven equation, but having the mass hierarchy as well as the
separation of different energy scales poses an extra complication to
the parameter setting and extrapolation as described above. We leave
such investigations for future works.

Of course, the stochastic method would not replace the diagrammatic
computations due to limited accuracy. Especially, for higher orders,
in addition to growing statistical uncertainties, the logarithmically
enhanced discretization may become an important limiting factor,
because they cannot be easily eliminated by numerical extrapolations.
Nevertheless, even with, for example, of order of ten percent
accuracy, it would provide useful estimate of the higher order
coefficients for which the diagrammatic calculation is difficult to
reach.

This work provides an evidence that the stochastic quantization method
is practically useful in the perturbative calculation of physical
quantities, and it may be more effective than the Feynman diagrams at
higher loops. 
We expect that the range of the applications to be much wider than QCD
and QED.
In general, once we encounter a problem which is sufficiently
complicated in the diagrammatic approach, the stochastic method may
provide an efficient means to approach.

\section*{Acknowledgements}
The work is supported by JSPS KAKENHI Grant Nos.~JP19H00689~(RK),
JP19K14711~(HT) and JP18H03710~(SH), and MEXT KAKENHI Grant
No.~JP18H05542~(RK).
This work is supported by the Particle, Nuclear and Astro Physics
Simulation Program No.~007 (FY2019) and No.~001 (FY2020) of Institute
of Particle and Nuclear Studies, High Energy Accelerator Research
Organization (KEK).

\appendix
\numberwithin{equation}{section}
\setcounter{equation}{0}

\begin{appendices}
\section{Ward-Takahashi identities}
\label{sec:WT}

The lattice action with $m_\gamma = 0$ is invariant under the
following BRS transformation:
\begin{align}
 \delta_{\rm BRS} \psi = i \lambda c \psi,
\end{align}
\begin{align}
 \delta_{\rm BRS} \bar \psi = -i \bar \psi \lambda c,
\end{align}
\begin{align}
 \delta_{\rm BRS} A_\mu = - {1 \over e} \lambda \nabla_\mu c,
\end{align}
\begin{align}
 \delta_{\rm BRS} \bar c = - {i \over e} \lambda B,
\end{align}
\begin{align}
 \delta_{\rm BRS} c = 0,
\end{align}
\begin{align}
 \delta_{\rm BRS} B = 0.
\end{align}
Here $\lambda$ is an anticommuting parameter. The fields $B$, $c$, and
$\bar c$ are the Nakanishi-Lautrup field, the ghost field, and the
anti-ghost field, respectively, with the action:
\begin{align}
 S_{B + c} = \sum_n 
\left[ \hat B(n) \nabla_\mu^* \hat A_\mu (n) - {\xi \over 2} \hat B(n)^2 
+i \hat{\bar c}(n) \nabla_\mu^* \nabla_\mu \hat c(n)
\right],
\end{align}
where the symbol ``$\ \hat{}\ $'' represents the fields multiplied by
the UV regulator, $\hat \Phi (n) = e^{-\nabla^2 / \Lambda_{\rm UV}^2}
\Phi (n)$.
The gauge fixing term is obtained by
integrating out the $B$ field. The equation of motion
\begin{align}
 B = {1 \over \xi} \nabla_\mu^* A_\mu,
\end{align}
holds as the operator equation, i.e.,
\begin{align}
 \langle B(n) {\cal O}_1 \cdots {\cal O}_m \rangle 
= {1 \over \xi} \langle \nabla_\mu^* A_\mu (n) {\cal O}_1 \cdots {\cal O}_m \rangle,
\end{align}
where ${\cal O}$'s are local operators which do not contain $B$. Also,
the ghosts are free fields in QED. Therefore,
\begin{align}
 \langle c(n) \bar c(m) \rangle =i \int {d^4 k\over (2 \pi)^4} 
 {e^{-2 \hat k^2 / \Lambda_{\rm UV}^2} \over \hat k^2} e^{i k \cdot (x_{n}-x_{m})}.
\end{align}

The BRS symmetry tells us
\begin{align}
\langle \delta_{\rm BRS} \left(
A_\mu (n_1) \bar c (n_2) \right) \rangle
= 0,
\end{align}
and 
\begin{align}
\langle \delta_{\rm BRS} \left(
\psi (n_1) \bar \psi (n_2) \bar c (n_3) \right) \rangle
 = 0.
\end{align}
These two equations lead the following Ward-Takahashi identities:
\begin{align}
 \hat k_\nu {\cal D}_{\mu \nu} (k)
= {\xi \hat k_\mu \over \hat k^2} e^{-2 \hat k^2 / \Lambda_{\rm UV}^2},
\label{eq:WT1}
\end{align}
\begin{align}
 \hat k_\mu G_\mu (p,k) 
= {e \xi \over \hat k^2} \left(
S (p+k) - S (p)
\right) e^{-2 \hat k^2 / \Lambda_{\rm UV}^2} .
\label{eq:WT2}
\end{align}
Here ${\cal D}_{\mu \nu}$, $S(p)$, and $G_\mu (p,k)$ are defined in
Eqs.~\eqref{eq:photonprop}, \eqref{eq:sf}, and \eqref{eq:gmu},
respectively.

The first identity~\eqref{eq:WT1} implies that the longitudinal part
of the propagator is exact at tree level, and thus
\begin{align}
\hat k_\mu {\cal D}^{-1}_{\mu \nu} (k)
= {\hat k^2 \over \xi} \hat k_\nu e^{2 \hat k^2 / \Lambda_{\rm UV}^2}.
\label{eq:WT1inv}
\end{align}
By using Eqs.~\eqref{eq:WT2}, \eqref{eq:WT1inv} and the definition in
Eq.~\eqref{eq:Gamma}, one finds
\begin{align}
 \hat k_\mu \left(-i e_{\rm P} \Gamma_\mu (p,k) \right)
= e \kappa \left(
S (p)^{-1} - S (p+k)^{-1}
\right).
\label{eq:vertexWT}
\end{align}

The Ward-Takahashi identities are modified when we include the mass term.
The identities are
\begin{align}
\langle \delta_{\rm BRS} \left(
A_\mu (n_1) \bar c (n_2) \right) \rangle
= \left \langle
\sum_{n_3} m_{\gamma}^2 \hat A_\nu (n_3) \left(
- { 1 \over e} \lambda \nabla_\nu \hat c (n_3) 
\right)
A_\mu (n_1) \bar c (n_2)
\right \rangle,
\end{align}
and 
\begin{align}
\langle \delta_{\rm BRS} \left(
\psi (n_1) \bar \psi (n_2) \bar c (n_3) \right) \rangle
= \left \langle
\sum_{n_4} m_{\gamma}^2 \hat A_\nu (n_4) \left(
- { 1 \over e} \lambda \nabla_\nu \hat c (n_4) 
\right)
\psi (n_1) \bar \psi (n_2) \bar c (n_3)
\right \rangle
.
\end{align}
These identities lead to
\begin{align}
 \hat k_\mu {\cal D}_{\mu \nu} (k)
= {\xi \hat k_\nu \over \hat k^2 + \xi m_{\gamma}^2} 
e^{-2 \hat k^2 / \Lambda_{\rm UV}^2},
\label{eq:modifiedWT1}
\end{align}
\begin{align}
 \hat k_\mu G_\mu (p,k) 
= {e \xi \over \hat k^2 + \xi m_\gamma^2} \left(
S (p+k) - S (p)
\right)
e^{-2 \hat k^2 / \Lambda_{\rm UV}^2}.
\end{align}
Again, Eq.~\eqref{eq:modifiedWT1} says that the longitudinal part of
the propagator is exact at the tree level. Together with the second
identity, one finds that Eq.~\eqref{eq:vertexWT} is maintained even
with the photon mass term.

\section{Numerical integration of the Langevin equation}
\label{sec:numerical_langevin}

We are going to solve the following type of equation,
\begin{align}
 \dot \phi (n,\tau) & = f(n,\tau) + \eta (n,\tau),
\end{align}
with the Gaussian noise,
\begin{align}
 \langle \eta (n,\tau) \eta (n', \tau') \rangle_\eta = 2 \delta_{nn'} \delta (\tau - \tau').
\end{align}
One can discretize the fictitious time $\tau$ as
\begin{align}
 \phi(n, \tau_{i+1}) - \phi(n, \tau_i) = f(n, \tau_i) \Delta \tau + \eta (n, \tau_i)
 \Delta \tau,
\end{align}
with
\begin{align}
 \langle \eta (n,\tau_i) \eta (n', \tau_j) \rangle_\eta = 2 \delta_{nn'} {1
 \over \Delta \tau} \delta_{ij}.
\end{align}
Now $\eta (n, \tau_i)$ are independent Gaussian noise with the
variance
\begin{align}
 \sigma^2 & = {2 \over \Delta \tau}.
\end{align}
Therefore, 
\begin{align}
 \phi(n, \tau_{i+1}) - \phi(n, \tau_i) = f(n, \tau_i) \Delta \tau + g(n, \tau_i)
 \sqrt{2 \Delta \tau},
\end{align}
where $g(n, \tau_i)$ are independent Gaussian noises with the variance
$\sigma^2 = 1$.
By taking the limit of small $\Delta \tau$, one can integrate the Langevin
equation.

A simple improvement is possible by using the Runge-Kutta type
algorithm~\cite{Helfand:1979}.
\begin{align}
 \phi(n, \tau_{i+1}) - \phi(n, \tau_i) = {1 \over 2} \left(
f_1 (n, \tau_i)
+ f_2 (n, \tau_i)
\right) \Delta \tau 
+ g(n, \tau_i)
 \sqrt{2 \Delta \tau},
\label{eq:RK}
\end{align}
where
\begin{align}
 f_1 (n, \tau_i) = f(n, \tau_i) \Big|_{\phi_i},
\end{align}
and
\begin{align}
 f_2 (n, \tau_i) = f(n, \tau_i) \Big|_{\phi_i + f_1 \Delta \tau + g \sqrt{2 \Delta \tau}}.
\label{eq:f2}
\end{align}
The Gaussian noises are common for Eqs.~\eqref{eq:RK} and
\eqref{eq:f2}.

\end{appendices}

\bibliography{bibcollection}
\bibliographystyle{arxiv-bibstyle/hyperieeetr2}

\end{document}